\documentclass{aa}  
\bibpunct{(}{)}{;}{a}{}{,} 
\usepackage{color}
\usepackage{graphicx}
\usepackage{txfonts}
\begin{document}

   \title{Low-temperature MIR to submillimeter mass absorption coefficient of interstellar dust analogues II:  Mg and Fe-rich amorphous silicates}


   \author{
        K. Demyk \inst{1}
 \and  C. Meny\inst{1}
 \and  H. Leroux\inst{2}
 \and  C. Depecker\inst{2}
 \and  J.-B. Brubach\inst{3}
 \and  P. Roy\inst{3}
 \and  C. Nayral\inst{4}
 \and   W.-S. Ojo\inst{4}
 \and  F. Delpech\inst{4}
   }
  
  \institute{
IRAP, Universit\'e de Toulouse, CNRS, UPS ; IRAP; 9 Av. colonel Roche, BP 44346, F-31028 Toulouse cedex 4, France
\and
UMET, UMR 8207, Universit\'e Lille 1, CNRS, F-59655 Villeneuve d'Ascq, France
\and
Ligne AILES - Synchrotron SOLEIL, L\'\ Orme des Merisiers, F-91192 Gif-sur-Yvette, France 
 \and
LPCNO, Universit\'e de Toulouse, CNRS, INSA, UPS, 135 avenue de Rangueil, 31077 Toulouse, France
   }

   \date{Received *** ; accepted ***}

 
  \abstract
   {To model the cold dust emission observed in the diffuse interstellar medium, in dense molecular clouds or in cold clumps that could eventually form new stars, it is mandatory to know the physical and spectroscopic properties of this dust and to understand its emission.}
   {This work is a continuation of previous studies aiming at providing astronomers with spectroscopic data of realistic cosmic dust analogues for the interpretation of observations. The aim of the present work is to extend the range of studied analogues to iron-rich silicate dust analogues.}
   {Ferromagnesium amorphous silicate dust analogues were produced by a sol-gel method with a mean composition close to $\mathrm{Mg_{1-x}Fe_{x}SiO_3}$ with x = 0.1, 0.2, 0.3, 0.4. Part of each sample was annealed at 500$^{\circ}$C for two hours in a reducing atmosphere to modify the oxidation state of iron. We have measured the mass absorption coefficient (MAC) of these eight ferromagnesium amorphous silicate dust analogues in the spectral domain 30 - 1000 $\mu$m for grain temperature in the range 10 - 300 K and at room temperature in the 5 - 40 $\mu$m range. }
   {The MAC of ferromagnesium samples behaves in the same way as the MAC of pure Mg-rich amorphous silicate samples. In the 30 - 300 K range, the MAC increases with increasing grain temperature whereas in the range 10 - 30 K, we do not see any change of the MAC. The MAC cannot be described by a single power law in ${\lambda}^{-\beta}$. The MAC of the samples does not show any clear trend with the iron content. However the annealing process has, on average, an effect on the MAC that we explain by the evolution of the structure of the samples induced by the processing. The MAC of all the samples is much higher than the MAC calculated by dust models. }
 {The complex behavior of the MAC of amorphous silicates with wavelength and temperature is observed whatever the exact silicate composition (Mg vs. Fe amount). It is a universal characteristic of amorphous materials, and therefore of amorphous cosmic silicates, that should be taken into account in astronomical modeling. The enhanced MAC of the measured samples compared to the MAC calculated for cosmic dust model implies that dust masses are overestimated by the models. }
 
\keywords{ Astrochemistry -  Methods: laboratory: solid state - Techniques: spectroscopic - (ISM:) dust, extinction - submillimeter: ISM- Infrared: ISM}

\titlerunning{Low temperature MIR/submm mass absorption coefficient of Fe-rich sol-gel silicates}
\authorrunning{K. Demyk et al.}
\maketitle
%

\section{Introduction}

The {\it Herschel} and {\it Planck} satellites have opened up the far infrared (FIR) and submillimeter (submm) spectral domain and we now have in hand a huge amount of observational data in the 250 $\mu$m - 1mm (850 - 300 GHz) domain. This is the domain where cold dust grains (10 - 30 K) emit and dominate the continuum emission. This FIR/submm emission traces cold astrophysical environments such as interstellar dense and diffuse clouds, cold clumps, and pre-stellar cores in our Galaxy as well as in external galaxies. It is used, for example, to derive the dust and cloud masses which constitute important information for star formation studies. Accurate knowledge and understanding of the dust emission is also important for cosmological studies requiring the subtraction of the foreground emission from our Galaxy. However a proper modeling of the FIR/submm dust emission is mandatory to making reliable interpretations of the observations. Dust emission is usually modeled using the Modified BlackBody model (MBB) and depends on the dust temperature and on the dust mass absorption coefficient (MAC) expressed as  $\mathrm{\kappa_{\lambda} =\:\kappa _{\lambda_{0} }\: ( \lambda {/} \lambda_{0}) ^{-\beta}}$ and characterized by a value at a reference wavelength, $\kappa _{\lambda_0}$, and by the spectral index, $\beta$, usually set to a value between 1 and 2, with no dependence on the temperature or wavelength. However, a great number of studies show that our understanding of cold dust emission is not complete. We refer the reader to \citet{demyk2017} for a detailed description and discussion of recent observational results. Briefly, it appears that dust-emission models are not able to fit the recent FIR/submm observations from the {\it Herschel} and {\it Planck} missions, independently of the level of noise in the observations, of the methods used to fit the data \citep{shetty2009b,juvela2012a,juvela2013}, and of temperature mixing along the line of sight \citep{malinen2011,juvela2012b}. These studies show that {\it {(i)}} the spectral index, $\beta$, is anti-correlated with the dust temperature \citep{planck2011_early_XXV,planck2011_early_XXIII,planck2013-XI,planck-XVII-Int-2014,juvela2015}, {\it {(ii)}} the $\beta$ value derived from the observations varies with the astrophysical environment \citep{galliano2011,paradis2014} and {\it {(iii)}} $\beta$ varies with the wavelength \citep{meisner2015,paradis2009}. These observational results may be understood in terms of variations of the dust nature (composition and structure) in various environments \citep{koehler2012,jones2013} or in terms of interaction of the dust with the electromagnetic radiation depending on the intrinsic dust physical properties \citep{meny2007}.  

With this study, our group continues the effort to investigate the optical properties of cosmic dust analogues in the mid infrared (MIR) to the millimeter domain as a function of temperature, undertaken 20 years ago by different groups \citep[we refer to][for the details of the studied samples]{demyk2013}. Briefly, \cite{agladze1996} were the first to study relevant interstellar silicate dust analogues in the temperature range from 1.2 K to 30 K and in the wavelength range from 700 $\mu$m to 2.9 mm. \cite{mennella1998} studied amorphous carbon samples and silicate samples in the 24 $\mu$m - 2 mm spectral domain and 24 - 300 K temperature range. \cite{boudet2005} investigated silica and silicate samples in the 10 - 300 K temperature range and in the spectral region 100 - 1000 $\mu$m. The work by \cite{coupeaud2011} was focussed on pure sol-gel Mg-rich silicates, of composition close to enstatite (MgSiO$_3$) and forsterite (Mg$_2$SiO$_4$), amorphous and crystalline, whose spectra were recorded in the 100 - 1000 $\mu$m spectral range and from 10 K to 300 K. These studies have brought important results about the spectroscopic characteristics and behavior of interstellar dust analogues in the FIR at varying temperature. They show that the MAC of the amorphous analogues (silicates and carbonaceous matter) increases with the grain temperature and that its spectral shape cannot be approximated with a single power law in the form ${\lambda}^{-\beta}$. The dependence of the MAC on the temperature, which is not observed in crystalline samples, is related to the amorphous nature and to the amount of defects in the structure of the material \citep{coupeaud2011}. A physical model was proposed by \cite{meny2007} to explain these experimental results. This model, named the TLS model, is based on a description of the amorphous structure of the material in terms of a temperature independent disordered charge distribution and of a collection of atomic configurations modeled as two-level systems and sensitive to the temperature. 

We extend this experimental work with the aim to deliver to the community a comprehensive and coherent data set measured on a common spectral domain (5 - 1000 $\mu$m) and temperature range (10 K - 300 K), on samples spanning a range of compositions, each set of which being synthesized using the same method. In \citet{coupeaud2011}, we investigated Mg-rich amorphous silicates synthesized with a sol-gel method. \citet{demyk2017} investigated four samples of amorphous Mg-rich glassy silicates with a composition close to enstatite, forsterite and one intermediate composition between forsterite and enstatite. 

The present study is focussed on ferromagnesium amorphous silicate dust analogues. \cite{mennella1998} derived, from transmission measurements, the MAC of one Fe-rich silicate amorphous sample, (Mg$_{0.18}$Fe$_{1.82}$SiO$_4$) in the MIR-mm domain (20 - 2000 $\mu$m) at low temperature (24 - 300 K). More recently, \cite{richey2013} measured the transmission and reflexion spectra of highly disordered, "chaotic", iron silicates (FeSiO) in the wavelength range 2 - 300 $\mu$m for grain temperature in the range 5 - 300 K. \cite{mohr2013} presented a preliminary study of a series of Mg/Fe-rich glassy pyroxene-like silicates of up to 50\% iron, in the spectral range 50 $\mu$m - 1.2 mm and from 300 K down to 10 K but they do not show nor discuss the low-temperature spectra in their article. Iron is highly depleted from the gas phase in the ISM. It is most probably incorporated into the cosmic dust grains although the form it takes in the grains remains poorly constrained. Iron could be present in the dust grains in the form of metallic iron or FeS inclusions, as can be observed in the presolar glass with embedded metal and sulfides (GEMS), \citet{bradley1994}) in porous interplanetary dust particles (IDPs). Based on elemental depletion observations and on the modeling of the dust formation, \cite{dwek2016} proposed that iron exists in metallic form either as inclusions or as separate iron grains. Iron has also been proposed to be present as a population of iron oxide grains (magnetite, Fe$_3$O$_4$,  or maghemite, $\gamma$-Fe$_2$O$_3$) \citep{draine2013a} or as iron oxide inclusions in the silicate grains. However, the amount of iron sulfides and iron oxides must be sufficiently low to be compatible with the absence of detection of the vibrational bands of these species. Iron could also be present in the amorphous silicate network either in the form of ferric iron (Fe$^{3+}$) and/or in the form of ferrous iron (Fe$^{2+}$). Interestingly, \cite{bose2012} and reference therein, showed that a number of presolar silicates are ferromagnesian. Such silicates thus constitute relevant analogues of interstellar silicates. In this article, we present the study of Mg-Fe-rich silicates produced by sol-gel method, of pyroxene mean composition close to (Mg$_{1-x}$Fe$_{x}$SiO$_3$) and containing iron from 10 to 40\% ($x$ = 0.1, 0.2, 0.3, 0.4). 

The paper is organized as follows: Sect.~\ref{exp} introduces the experimental procedures for the dust analogues synthesis, their characterization, and the spectroscopic measurements, Sect.~\ref{spectro} presents the MAC of the studied samples measured in the 5 - 1000 $\mu$m range and in the 10 - 300 K temperature range and Sect.~\ref{discussion} discusses these results and their implications for astrophysical studies.


\section{Experiments}
\label{exp}

\subsection{Sample synthesis}

Amorphous samples of composition Mg$_{1-x}$Fe$_x$SiO$_3$, with $x$ = 0.1, 0.2, 0.3, 0.4 were synthesized with a sol-gel method using nitrates as precursors. The method is described in detail in \citet{gillot2009}. In the absence of iron, a clear and transparent gel is formed. The gel is more and more brown as the Fe concentration of the final product increases. When the iron nitrate precursor is used, it is necessary to adjust the pH ($\sim$1.7) with ammonium hydroxide in order to allow a reasonable gelification time and to avoid the precipitation of iron hydroxide. Once gelification has been reached, the gel is aged at ambient temperature for fifteen minutes before being dried at 110$^{\circ}$C for 48 hours in an oven under primary vacuum. At this stage, the translucent gel completely expanded in the liquid shrinks dramatically by losing around 50-75 \% of its volume. Finally, the dried gel, called xerogel, is ground in an agate mortar before the purification stage at 500$^{\circ}$C, in air, for two hours. Using these conditions for the synthesis, ferric iron (Fe$^{3+}$ or FeIII) is dominant. These samples are named E10 to E40 for samples containing 10 to 40\% iron. The ferric iron was partially reduced to ferrous iron (Fe$^{2+}$ or FeII) by loading the dehydrated gel into an oven with a gas streaming (Ar + 10\% H$_2$) at 500$^{\circ}$C for 3 hours. These samples, in which part of the iron has been reduced, are named E10R to E40R  for samples containing 10 to 40\% iron; we  also refer to these later in the article as "processed" samples. \\

\subsection{Sample characterization}
\label{carac}

\begin{figure}[!t]
\begin{center}
\includegraphics[scale=1]{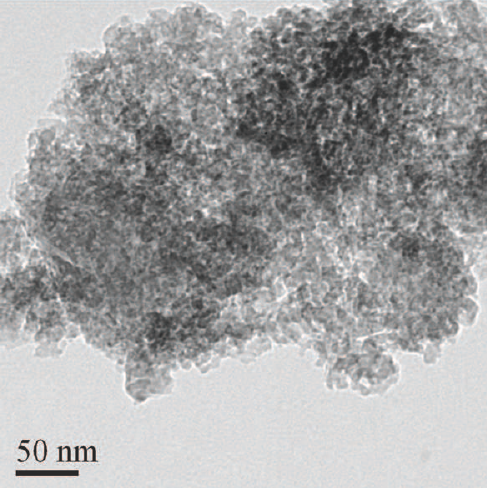}
         \caption{TEM image of one sample. The sol-gel samples are constituted by clusters about 10 nm in size. }
             \label{Fig1_TEM}
\end{center}
\end{figure}

\begin{figure}[!t]
\includegraphics[scale=.8]{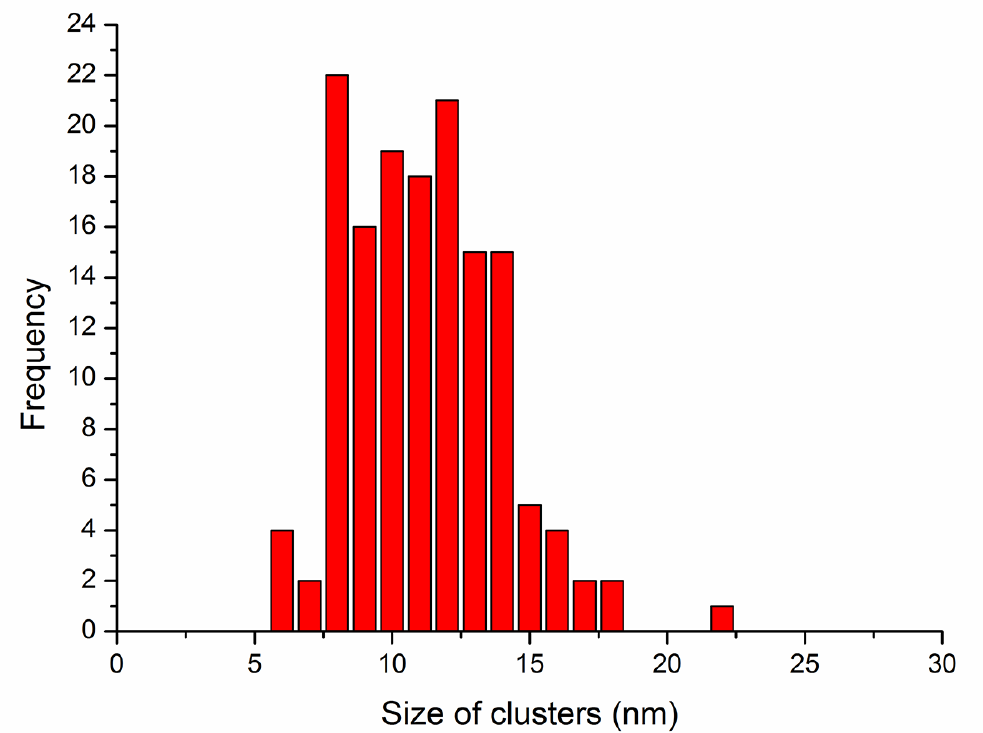}
         \caption{Size distribution of the clusters constituting the sol-gel samples.}
             \label{Fig2_size}
\end{figure}

\begin{figure}[!h]
\begin{center}
\includegraphics[scale=.73, trim={0.0cm 0.0cm 0.0cm 0.0cm},clip]{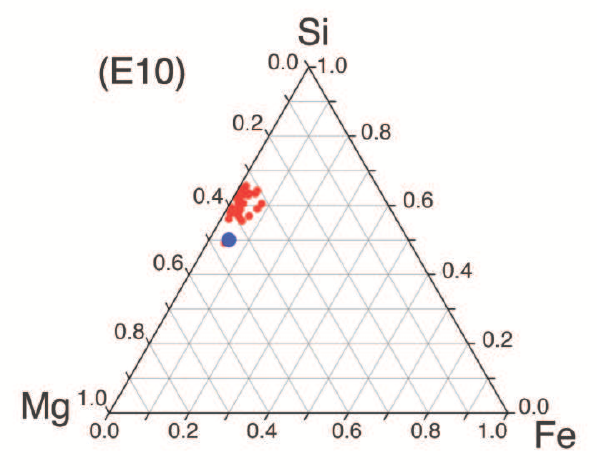}
\includegraphics[scale=.73, trim={0.0cm 0.0cm 0.0cm 0.0cm},clip]{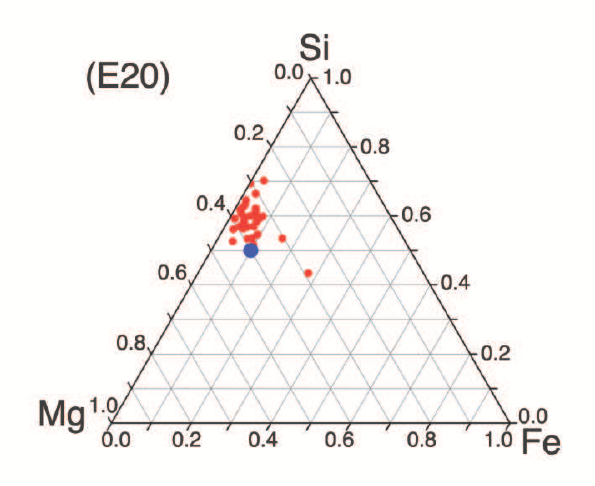}\\
\includegraphics[scale=.65, trim={0cm 0cm 0cm 0cm},clip]{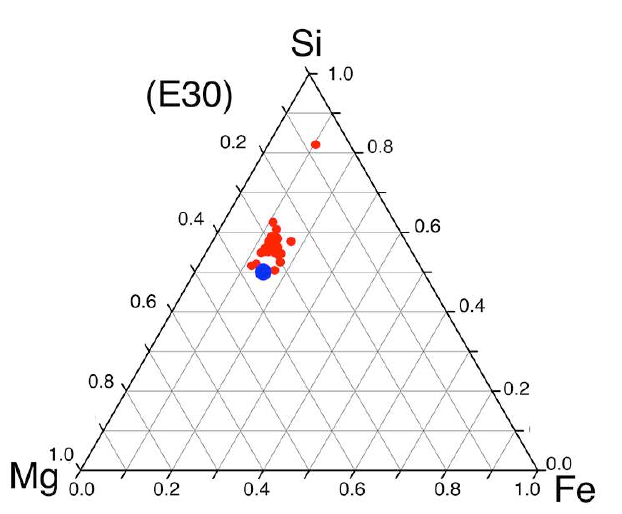}
\includegraphics[scale=.65, trim={0cm 0cm 0cm 0cm},clip]{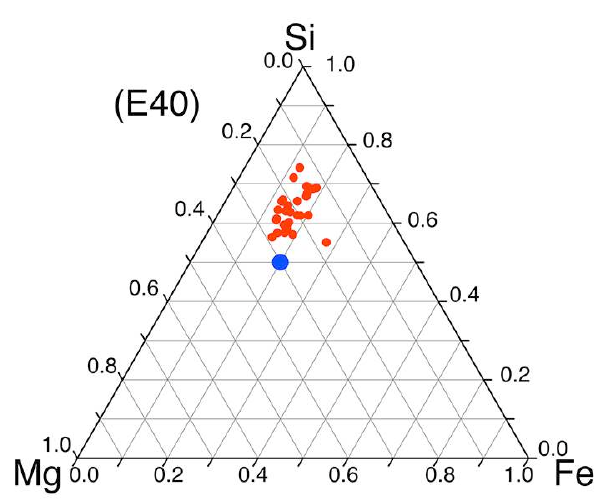}
         \caption{Fe-Mg-Si ternary diagram (at \%) showing the composition field of the sol-gel samples, E10 (top left panel), E20 (top right panel), E30 (bottom left panel) and E40 (bottom right panel). Each red point represents an individual analysis with a selected volume of matter of $\sim$ 10$^{-3}$ $\mu$m$^3$. The blue point is the target composition. The slight offset between the target composition and the sol-gel sample is likely due to a small Mg loss under the beam during the analyses.}
             \label{Fig3_compo}
             \end{center}
\end{figure}

The sol-gel synthesis products were characterized by transmission electron microscopy (TEM) in order to infer their morphology at the nanometer scale and their local chemical composition. The TEM examination requires thin samples (typically less than 100 nm thick). To prepare the samples, pieces of sol-gel blocks were crushed in alcohol. A drop of alcohol, containing a large number of small fragments in suspension, was deposited on a carbon film supported by a TEM copper grid. The TEM characterization was performed using a FEI Tecnai G2-20 twin at the electron microscopy facility of the University of Lille. The sample morphology and size distribution were studied by conventional bright field imaging. It is similar to the Mg-rich sol-gel samples from \cite{coupeaud2011}. Whatever their composition (x = 0, 0.1, 0.2, 0.3, 0.4), the samples consist of clusters of matter bonded to one another (Fig.~\ref{Fig1_TEM}). In all samples, the clusters are homogenous in size, centred around 11 nm with a Gaussian size distribution between 5 and 20 nm (Fig.~\ref{Fig2_size}). The size of the porosity is of the same order of magnitude ($\sim$ 10 nm). The amorphous state of the clusters is confirmed by electron diffraction patterns which show diffuse rings characteristic of amorphous matter. Compositions were measured by EDS by selecting volumes typically of the order of 10$^{-3}$ $\mu$m$^3$, thus including a large number of clusters in each analysis. The use of smaller volumes was avoided because it led to significant damage of the samples under the electron beam and a preferential loss of Mg with respect to Si and Fe. The measured compositions are found to be relatively homogeneous and reasonably close to the target compositions (Fig.~\ref{Fig3_compo}), although a slight deficit of Mg was systematically observed, likely due to degradation of the samples under the electron beam. The E10 and E20 samples appear to have very similar measured composition, as do the samples E30 and E40 (Fig.~\ref{Fig3_compo}). 

We investigated the oxidation state of the iron contained in the samples with M\"ossbauer spectroscopy. M\"ossbauer spectra were collected on a constant-acceleration conventional spectrometer with a 1.85 GBq source of $^{57}$Co (Rh matrix) at 293 K. The absorber was a sample of ca. 50 - 100 mg of powder that was enclosed in a 20-mm-diameter cylindrical plastic sample holder, the size of which has been determined to optimize the absorption. 
We observe doublets and sextets in the M\"ossbauer spectra of the samples which are characteristic of quadrupole and magnetic dipole interactions, respectively, the doublets being associated to the presence of iron in the silicate network and the sextets to oxide phases with magnetic signature \citep{ferreiradasilva1992,jayasuriya2004}. The M\"ossbauer spectra of samples E10 to E40 show a single doublet characteristic of FeIII in silicates whereas the spectra of the E10R to E40R samples show two doublets characteristic of FeIII and FeII. This indicates that the annealing of the samples under reducing conditions (Ar + 10\% H$_2$) led to incomplete iron reduction. It is likely due to the low temperature of annealing (500$^{\circ}$C). Sol-gel processing at higher temperature is precluded because of the strong propensity of the samples to crystallization. We also observed sextets in the spectra that reveal the presence of oxides. In the unprocessed samples, two sextets are observed (only one sextet is observed for the E10 sample) and their isomer shifts point to disordered (and/or small nanocrystals of) hematite (Fe$_2$O$_3$, FeIII) as a carrier \citep{ferreiradasilva1992}. Two sextets are also observed for the E30R and E40R samples, with a relative intensity of $\sim$ 2:1 but their isomer shifts are different from those of the sextets observed in the spectra of the unprocessed samples suggesting that the hematite has been reduced into magnetite (Fe$_3$O$_4$, FeII:FeIII = 1:2) \citep{lyubutin2009}. The measured isomer shift domain was not extended enough to measure the sextet for the E10R and E20R samples for which we have no information. From the M\"ossbauer spectra, we estimate that the samples contained about 5 - 10\% iron oxide, which means that the main fraction of iron is present within the silicate sol-gel. 

\subsection{Spectroscopic measurements}
\label{spectro}

The spectroscopic measurements were performed on the setup ESPOIRS at IRAP in the spectral range 5 - 1000 $\mu$m and on the AILES beam line at the synchrotron SOLEIL in the spectral range 250 - 1000/1200 $\mu$m. The ESPOIRS setup is dedicated to the characterization of interstellar dust analogue spectroscopic properties. Thanks to a set of detectors, beamsplitters, and sources it covers the spectral domain from 0.7 $\mu$m to $\sim$ 1000 $\mu$m. In the FIR ($\lambda$ $\ge$ 30 $\mu$m), we use a TES Si bolometer detector from QMC Instrument (operating at 8 K and cooled with a He pulse-tube), a silicon beamsplitter, and a mercury lamp. In the MIR, we use a CsI beamsplitter, a Globar source and a DLaTGS detector. The samples are cooled down to 10 K with a pulse tube cooled cryostat. The AILES beam line is equipped with a similar experimental setup described in detail by \citet{brubach2010}. In the MIR range (5 - 40 $\mu$m), the transmission spectra were measured at room temperature whereas in the FIR/submm range ($\lambda$ $\ge$ 30 $\mu$m) they were measured at 10, 30, 100, 200 and 300 K.

The samples are prepared for transmission measurements in the form of pellets of 13 mm diameter. In the MIR, KBr (Aldrich) pellets are pressed at room temperature under ten tons for several minutes. In the FIR, polyethylene (PE, Thermo Fisher Scientific) pellets are pressed under ten tons after annealing of the PE+sample mixture at 130$^{\circ}$C for five minutes. To obtain the MAC of the samples in the full wavelength range from 5 $\mu$m to 1 mm, several pellets are made with increasing mass of sample to compensate for the decrease of the MAC of the sample with increasing wavelength. Typically, 0.5 mg of sample is enough to measure the MIR spectrum whereas more than 100 mg of sample is required for measurements around 1 mm \citep[we refer to][for more details]{demyk2017}. 

The final MAC curves are constructed from the MAC curves of the pellets containing different masses of sample. The effect of the KBr and PE matrix is taken into account following the procedure described in \cite{mennella1998}. The error on the measured MAC is the quadratic sum of the uncertainty on the thermal stability of the spectrometer and on the uncertainty on the mass of sample in the pellet. The analysis of the spectral data to reconstruct the MAC of the samples and the details on the error determination are explained in  \citet{demyk2017}.


\begin{figure}[!t]
\includegraphics[scale=.45,trim={1cm 1cm 1cm 1cm},clip]{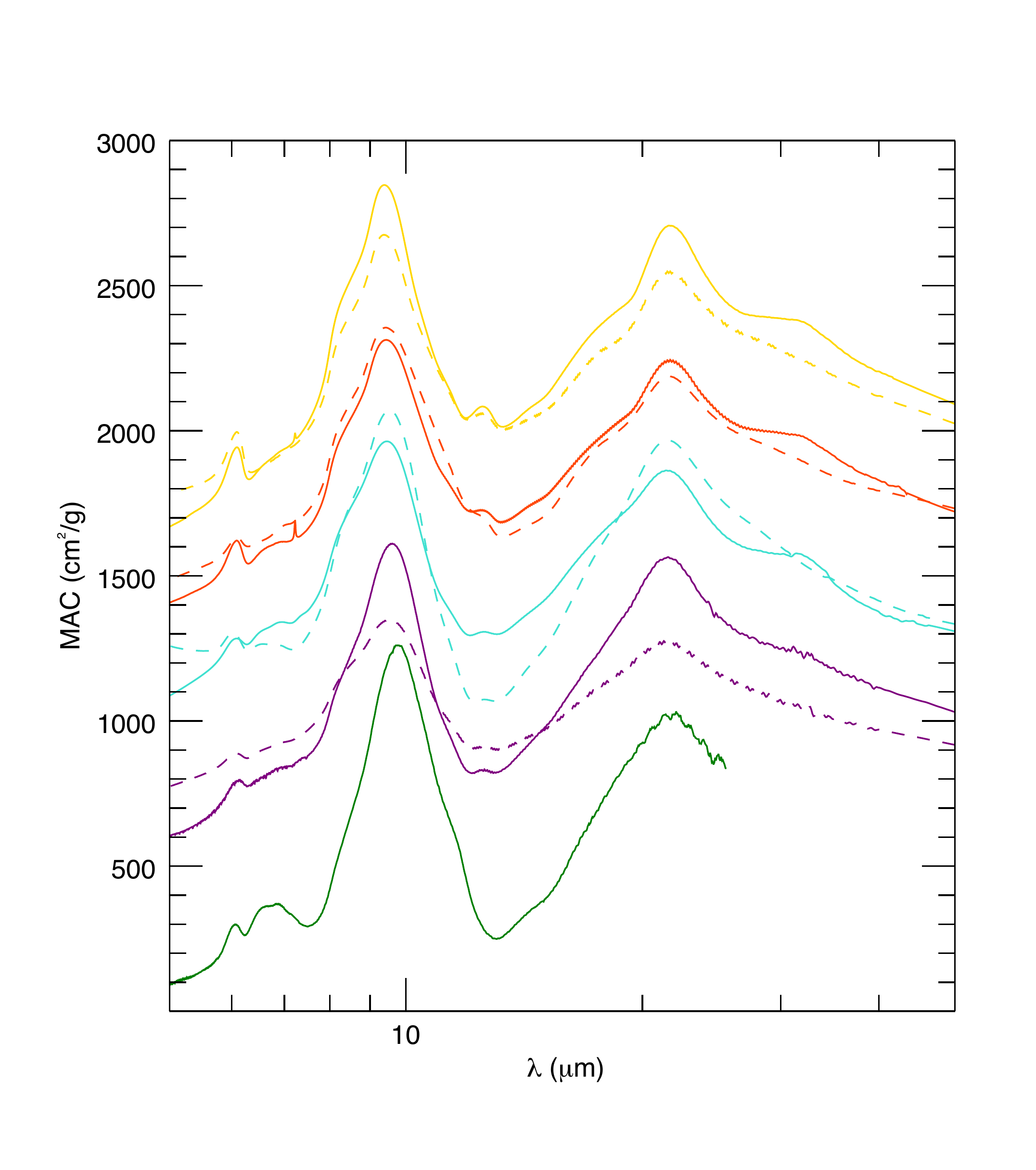}
         \caption{MAC of the samples in the MIR domain at room temperature. The continuous lines designate the samples E10 (purple), E20 (turquoise), E30 (red) and E40 (yellow) while the dashed lines designate the E10R, E20R, E30R and E40R samples (same color). The green curve shows the MAC of the E00 sample from \cite{coupeaud2011}. For the sake of clarity, the MAC curves of the samples E10, E20, E30, E40, E10R, E20R, E30R and E40R have been shifted.}
             \label{kappamir}
\end{figure}

\section{MIR  and FIR/submm mass absorption coefficient as a function of temperature}
\label{spectro}

The MAC of all samples, measured at room temperature in the range 5 - 40 $\mu$m, are shown in Fig.~\ref{kappamir}, together with the MAC of the MgSiO$_3$ sample from \citet{coupeaud2011} (named E in \citet{coupeaud2011} and hereafter named E00). This sample, which was synthesized with the same method as the samples studied here, is considered in this study because it represents the iron-free counterpart of the Fe-rich samples. Figure~\ref{kappamir} shows the stretching vibration of the Si-O bond of the SiO$_4$ tetrahedra at $\sim$ 9.6 $\mu$m and the bending vibration of the Si-O-Si bond at $\sim$ 21.6 $\mu$m. The peak position and shape of the vibrational bands of silicates are influenced by the structure of the material, which can be described by the relative amount of non-bringing oxygen per tetrahedra (NBO/T, non-bringing oxygen atoms are oxygen atoms which belong to a single tetrahedra). They are also influenced by the presence and nature of the cations within the silicates. For Mg- and Fe-rich silicates, it is known that the position of the stretching band is shifted toward short wavelengths when the Fe content increases \citep{dorschner1995}. As seen from Table~\ref{MIR}, this is verified by our two sets of samples. The peak position of the stretching vibration is 9.61 $\mu$m for sample E10 and it decreases to 9.38 $\mu$m for sample E40. The peak position is 9.56 and 9.37 $\mu$m for sample E10R and E40R, respectively. For comparison, it is 9.76 $\mu$m for sample E00. For a given iron content, the stretching modes of the unprocessed samples (Exx) and of the processed samples (ExxR) do not present strong differences in terms of peak position of the band nor in terms of width. The slight asymmetry of the stretching mode, which exhibits a shoulder at $\sim$ 8.3 $\mu$m, together with the weak band at 12.5 $\mu$m, might indicate the presence of some silica (SiO$_2$) within the samples. However it is not possible from the MIR spectra to be more specific regarding the amount of silica, its structure and the size of the inclusions within the silicate matrix. The presence of silica-rich material is likely associated with the formation of iron-oxide. The E00 sample does not present the same spectral feature and should not contain silica.

\begin{table}[!t]
\caption{Peak position of the vibrational bands observed in the MIR domain. }
\label{MIR}
\begin{center}
\begin{tabular}{c c c}
\hline
\hline
        &       \multicolumn{2}{c}{Peak Position ($\mu$m)}       \\
Samples &       Stretching mode  & Bending mode  \\
\hline
E00     &       9.76 & 22.03 \\
E10     &       9.61 & 21.56 \\
E20     &       9.50 & 21.56 \\
E30     &       9.40 & 21.69 \\
\vspace{0.1cm}
E40     &       9.38 & 21.60 \\
E10R    &       9.56 & 21.56 \\ 
E20R    &       9.53 & 21.47 \\
E30R    &       9.43 & 21.60 \\
E40R    &       9.37 & 21.51 \\
\hline
\hline
\end{tabular}
\end{center}
\label{default}
\end{table}%

\begin{table*}[!t]
\caption {Value of the MAC of the E10, E20, E30, E40, E10R, E20R, E30R and E40R samples and for cosmic dust models.} 
\label{table_kappa}
\begin{center}
\begin{tabular}{llc|cc|cc|cc|cc}
\hline 
\hline 
  & \multicolumn{10}{c}{Mass Absorption Coefficient (cm$^2$.g$^{-1}$)}  \\
\cline{2-11} &  \multicolumn{2}{c}{100 $\mu$m}   &  \multicolumn{2}{c}{250 $\mu$m}  &  \multicolumn{2}{c}{500 $\mu$m}      & \multicolumn{2}{c}{850 $\mu$m} &    \multicolumn{2}{c}{1 mm}   \\ 
& 10K & 300K            &  10K & 300K           & 10K & 300K            & 10K & 300K              & 10K & 300K     \\ 
\cline{2-11}    E10     & 260.0 & 305.3         & 37.7 & 92.6   & 7.2 & 20.6         & 1.0 & 6.0     & -     & -     \\
E20     & 224.8 & 243.5         & 46.2 & 58.8   & 6.8 & 14.2    & 1.5 & 4.9         & 1.0 &  3.7 \\
E30     & 195.2 & 217.8         & 41.0 & 52.9   & 7.6 & 14.1    & 2.8 & 6.0         & 2.4 & 5.1  \\
\vspace{0.2cm}
E40     & 219.3 & 245.0         & 35.7 & 62.5   & 5.8 & 15.7    & 2.2 & 7.0         & 2.0 & 6.1  \\
E10R &  244.0 & 270.4   & 57.5 & 80.0   & 11.5 & 23.6   & 3.9 & 10.4    & 3.5 & 8.9   \\
E20R & 249.4 & 268.6    & 55.1 & 79.2   & 14.4 & 24.6   & 7.0 & 12.2    & 5.8 & 10.1 \\
E30R & 225.7 & 253.9    & 44.7 & 67.3   & 6.8   & 18.8  & 1.4 & 8.0     & 0.8 & 6.2   \\
\vspace{0.2cm}
E40R & 195.2 & 219.0    & 39.2 & 53.7   & 5.0   & 15.2   & 1.3 & 7.4    & 1.0 & 6.1   \\
 < MAC$\mathrm{_{Exx}}$>\tablefootmark{(1)} & 225 & 253 &       40 & 67 &       6.8 & 16.1 & 1.9 & 6.0 & 1.4 & 4.8    \\
 < MAC$\mathrm{_{ExxR}}$>\tablefootmark{(2)} & 229 & 253 & 49 & 70 &    9.4 & 20.6 & 3.4 & 9.5 & 2.8 & 7.8    \\
\vspace{0.2cm}
 < MAC$\mathrm{_{all}}$>\tablefootmark{(3)} & 227       & 253 &         45 & 68 &  8.1 & 18.3 & 2.6 & 7.7 & 2.1& 6.3         \\
MAC sphere 0.1  $\mu$m  \tablefootmark{(4)}& \multicolumn{2}{c|}{33.5} & \multicolumn{2}{c|}{5.1} & \multicolumn{2}{c|}{1.2} & \multicolumn{2}{c|}{0.48} & \multicolumn{2}{c}{0.36}\\
MAC distrib spherical grains \tablefootmark{(4)} & \multicolumn{2}{c|}{36.9} & \multicolumn{2}{c|}{4.9} & \multicolumn{2}{c|}{1.1} & \multicolumn{2}{c|}{0.45} & \multicolumn{2}{c}{0.34}\\
MAC distrib prolate grains \tablefootmark{(4)} & \multicolumn{2}{c|}{50.3} & \multicolumn{2}{c|}{6.8} & \multicolumn{2}{c|}{1.5} & \multicolumn{2}{c|}{0.6} & \multicolumn{2}{c}{0.47}\\
\vspace{0.2cm}
MAC CDE \tablefootmark{(4)} & \multicolumn{2}{c|}{74.0} & \multicolumn{2}{c|}{11.5} & \multicolumn{2}{c|}{2.6} & \multicolumn{2}{c|}{1.05} & \multicolumn{2}{c}{0.81}\\
\hline
\hline
\end{tabular} 
\tablefoot{ \\
\tablefoottext{1}{MAC averaged over the four unprocessed samples E10, E20, E30, E40.} \tablefoottext{2}{MAC averaged over the four processed samples E10R, E20R, E30R, E40R.} \tablefoottext{3}{MAC averaged over the eight samples E10, E20, E30, E40 and E10R, E20R, E30R, E40R.}
\tablefoottext{4}{These MAC are calculated using the optical constants of the "astrosilicates" from \citet{li2001} (Sect.~\ref{dust_model}).\\} 
}
\end{center}
\end{table*}

The peak position of the bending band does not seem to follow a trend with the iron content. It varies only slightly in the ranges 21.56 - 21.69 $\mu$m and  21.56 - 21.60 $\mu$m for the Exx and ExxR samples, respectively (22.03 $\mu$m for sample E00). The band exhibits a shoulder peaking at $\sim$ 17 $\mu$m for all samples (unprocessed and processed) which is more and more pronounced as the iron fraction increases. It is therefore probably related to an iron phase which is not affected by the processing applied to the samples. 
We note that the position of the shoulder is close to the position of the vibrational stretching band calculated for small spherical grains of Fe$_x$Mg$_{1-x}$O oxides \citep{henning1995}. The main difference between the bending mode of the unprocessed and processed samples is the presence of a band at $\sim$ 32.3 $\mu$m in the spectra of the unprocessed samples but not in the one of the processed samples. The intensity of this band increases with the amount of iron, therefore it is probably related to iron in a form which is altered by the processing. It could be iron oxides but it is difficult to identify which oxides, since they are most likely amorphous and of very small size. The peak position of this band is in agreement with the transmission spectrum of fine particles of hematite measured by \cite{marra2011}, the other features characteristic of hematite being hidden by the silicate bands. The fact that this band is not seen in the spectra of the processed samples indicates that the reduction process has operated even though the M\"ossbauer spectra show that it is not complete (see Sect.~\ref{carac}). The processed samples might contain some iron oxides but in small amounts since no IR spectral feature may be assigned to them. To summarize, the analysis of the MIR spectra suggests that the samples are not chemically homogeneous at small scales ($\le$ 100 nm, see Sect.~\ref{carac}). They might contain some three-dimensional structures compatible with SiO$_2$ and some iron oxide phases which should be different in the unprocessed and processed samples. \\

\begin{figure*}[!th]
\begin{center}
\includegraphics[scale=.29,trim={0cm 1cm 0cm 1cm},clip]{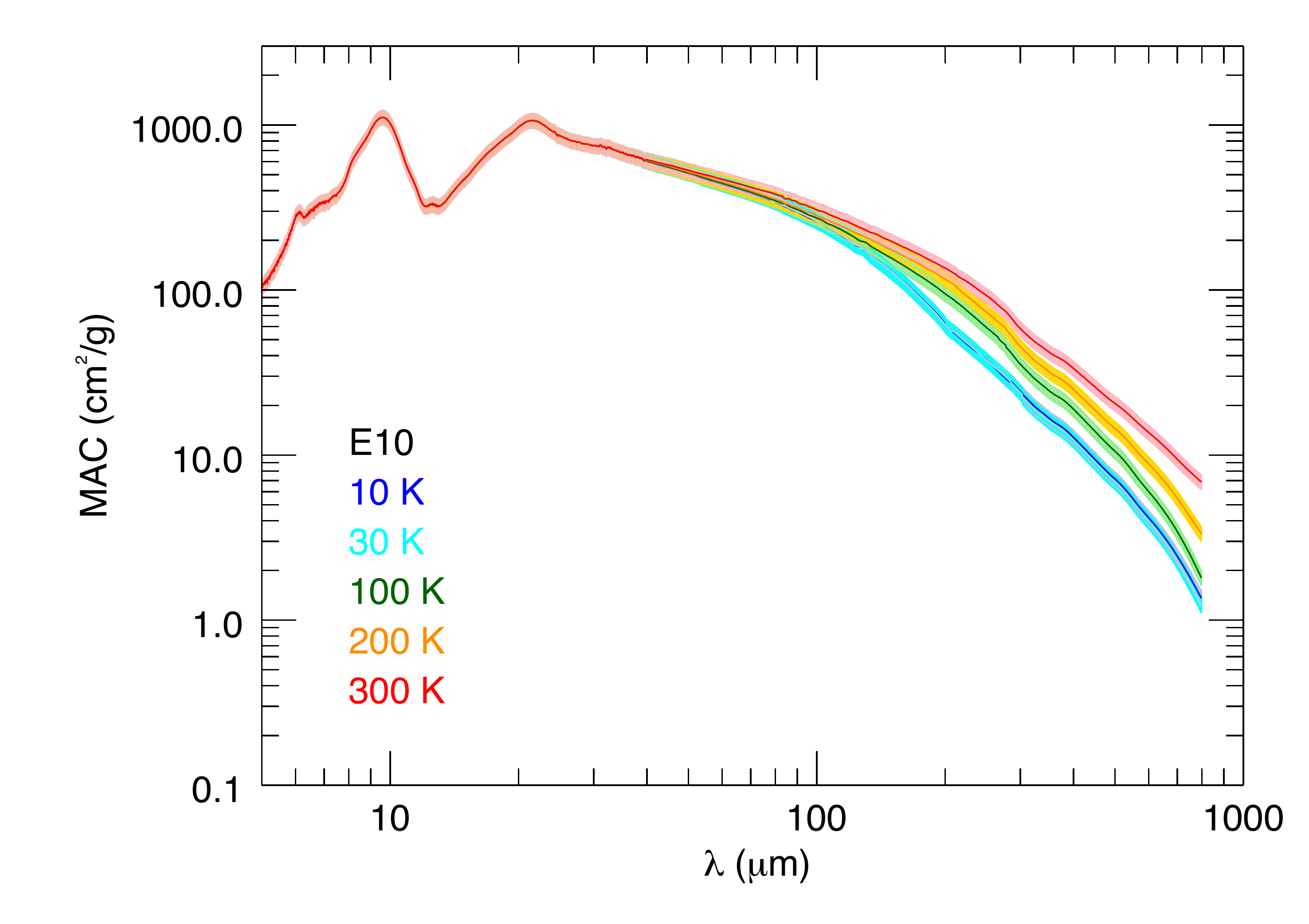}
\includegraphics[scale=.29,trim={0cm 1cm 0cm 1cm},clip]{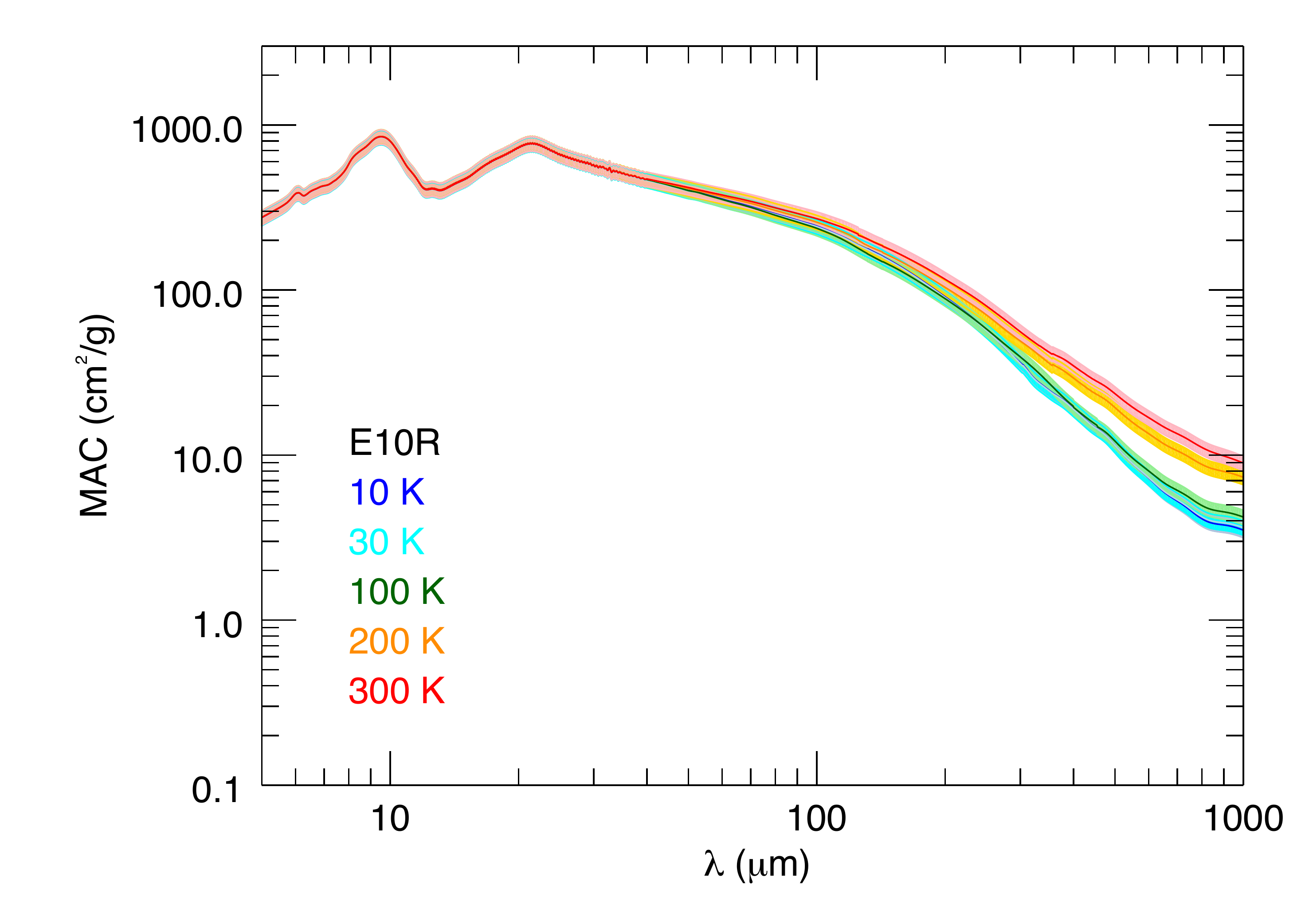}\\
\includegraphics[scale=.29,trim={0cm 1cm 0cm 1cm},clip]{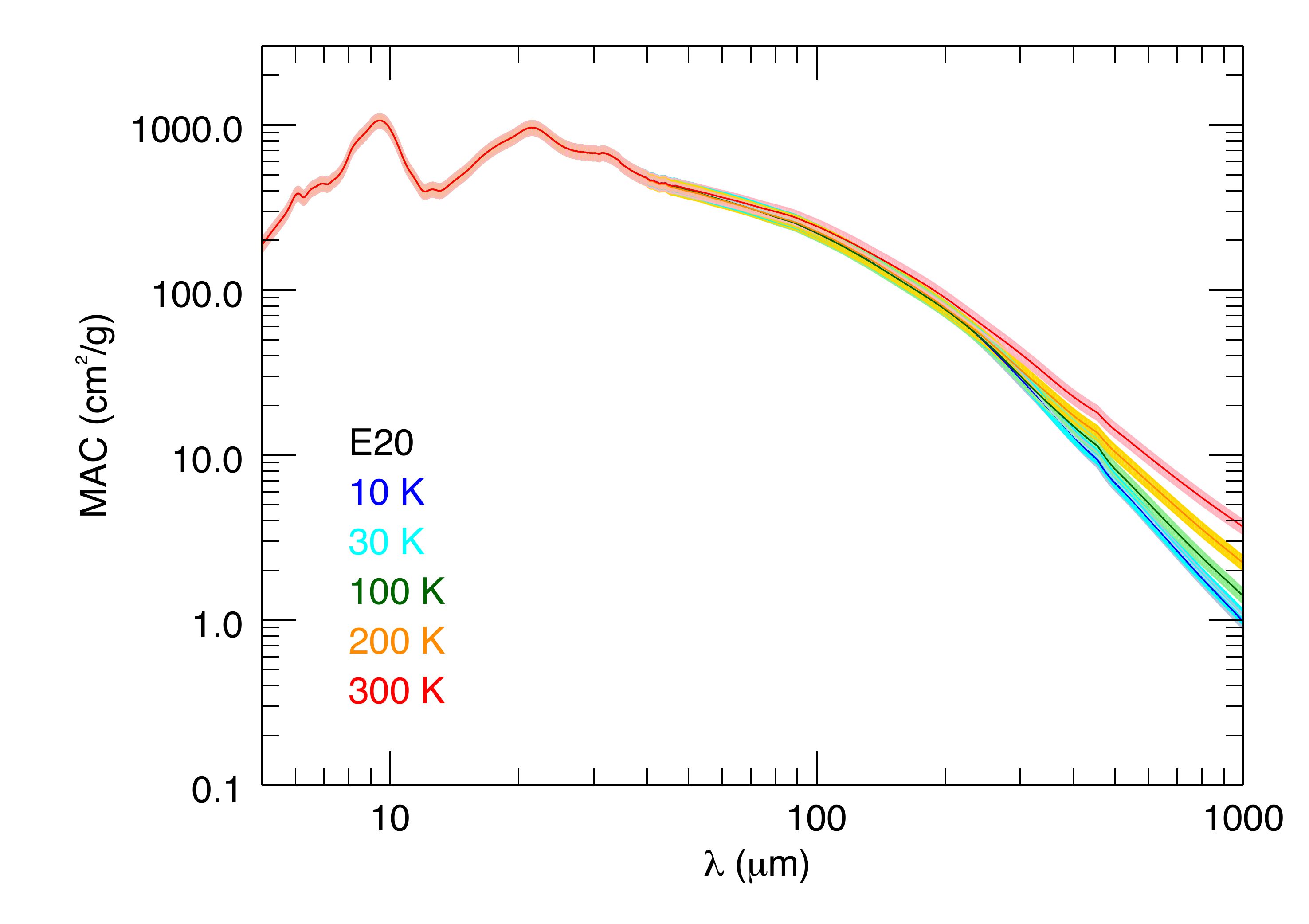}
\includegraphics[scale=.29,trim={0cm 1cm 0cm 1cm},clip]{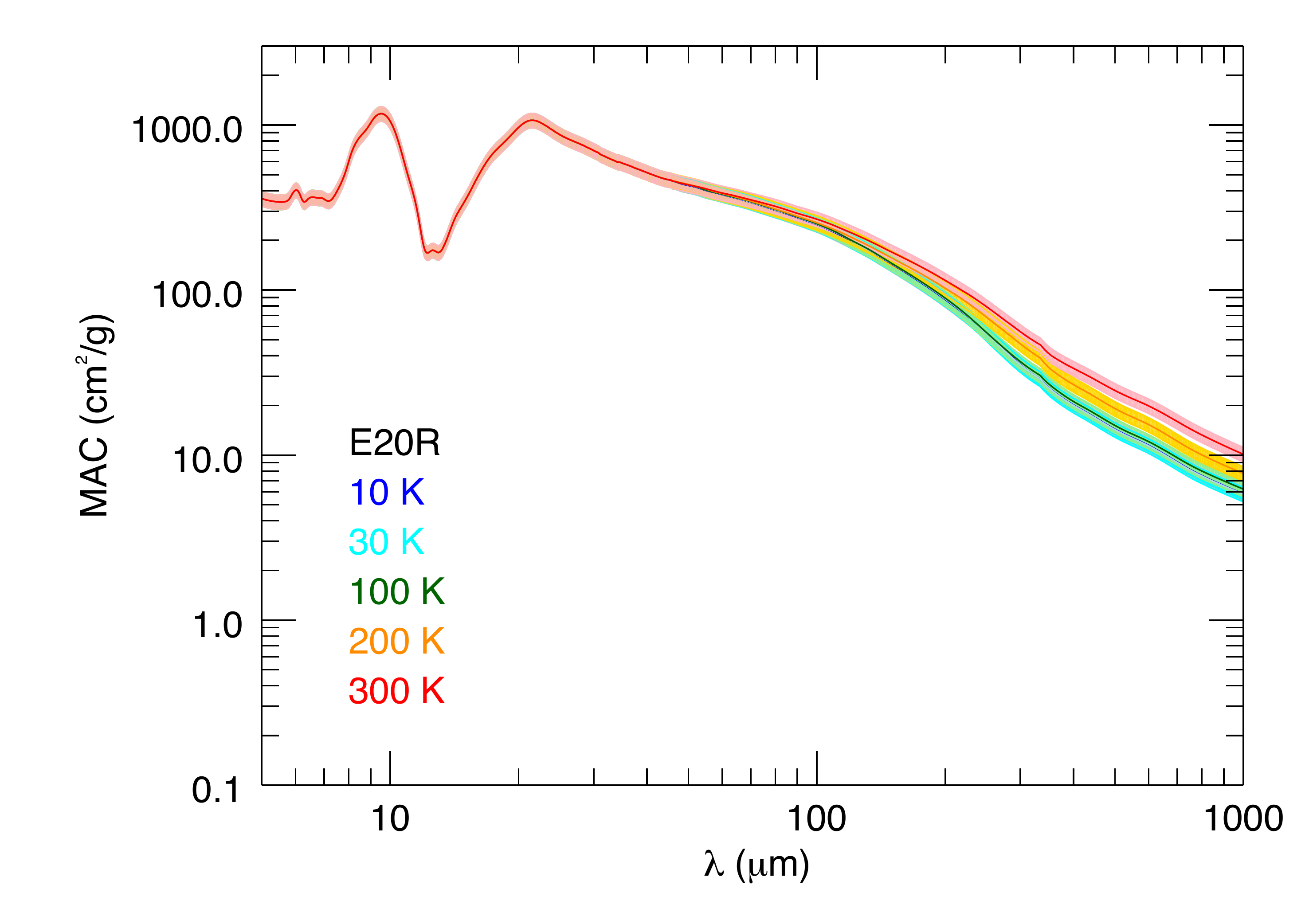}\\
\includegraphics[scale=.29,trim={0cm 1cm 0cm 1cm},clip]{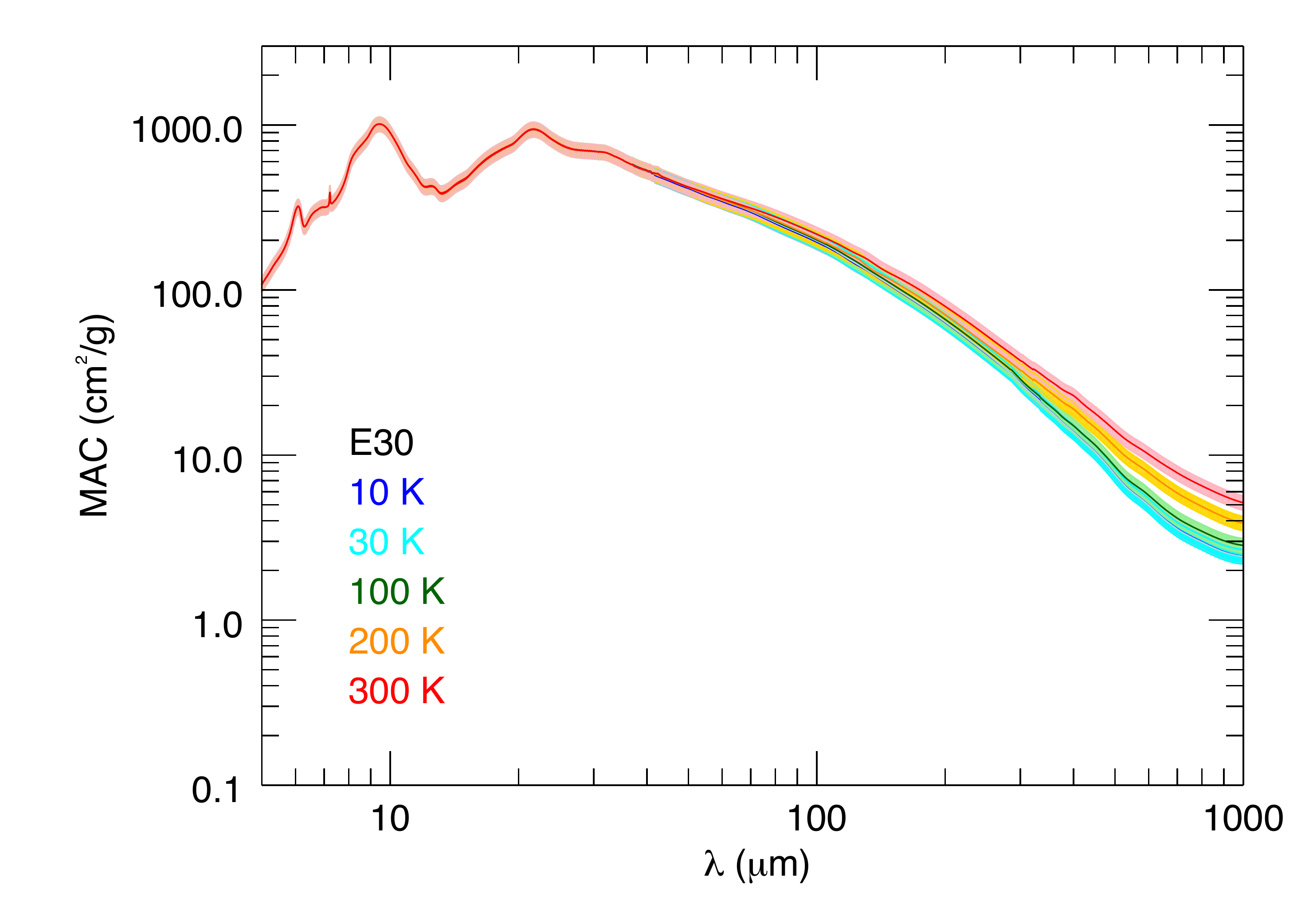}
\includegraphics[scale=.29,trim={0cm 1cm 0cm 1cm},clip]{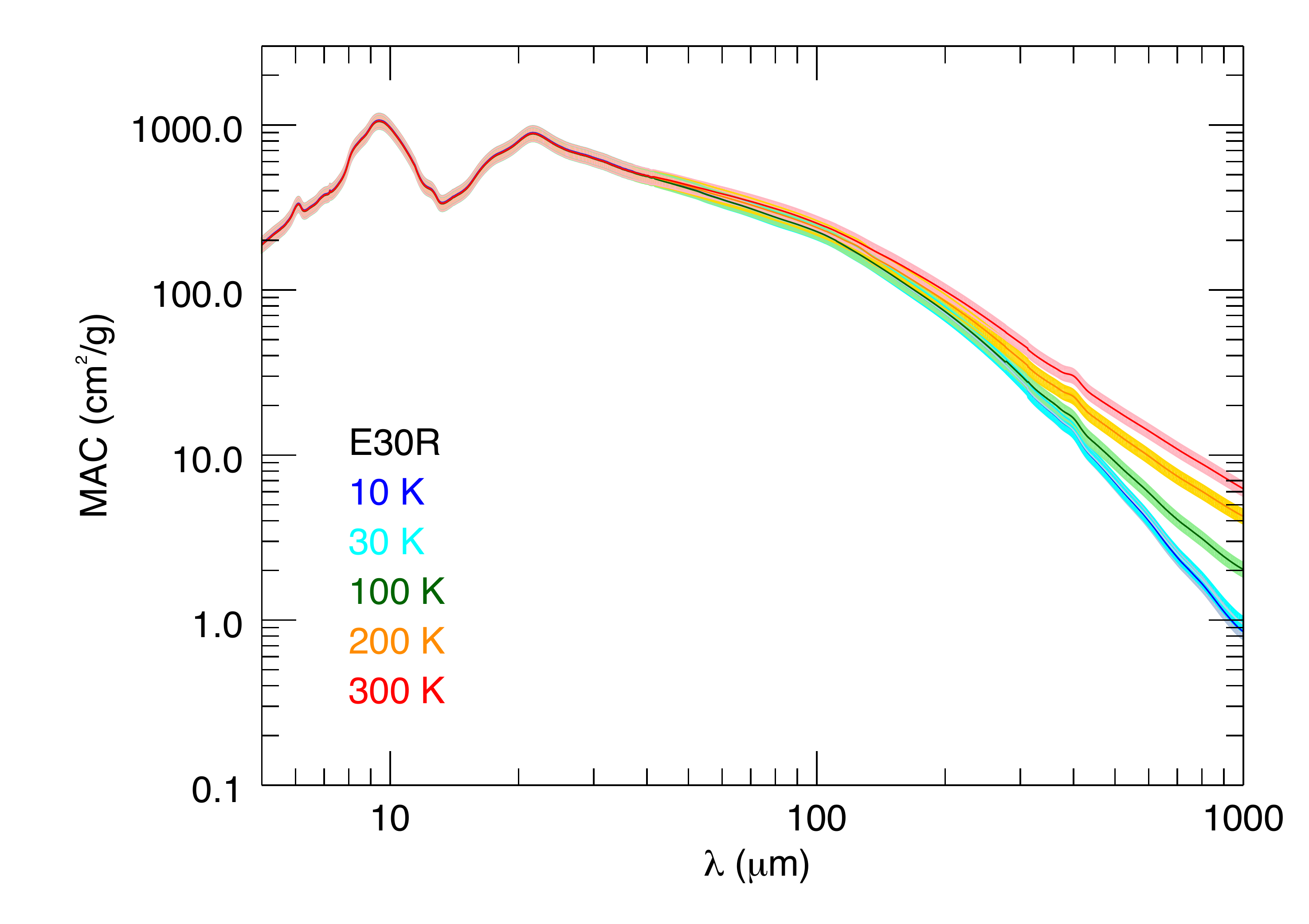}\\
\includegraphics[scale=.29,trim={0cm 1cm 0cm 1cm},clip]{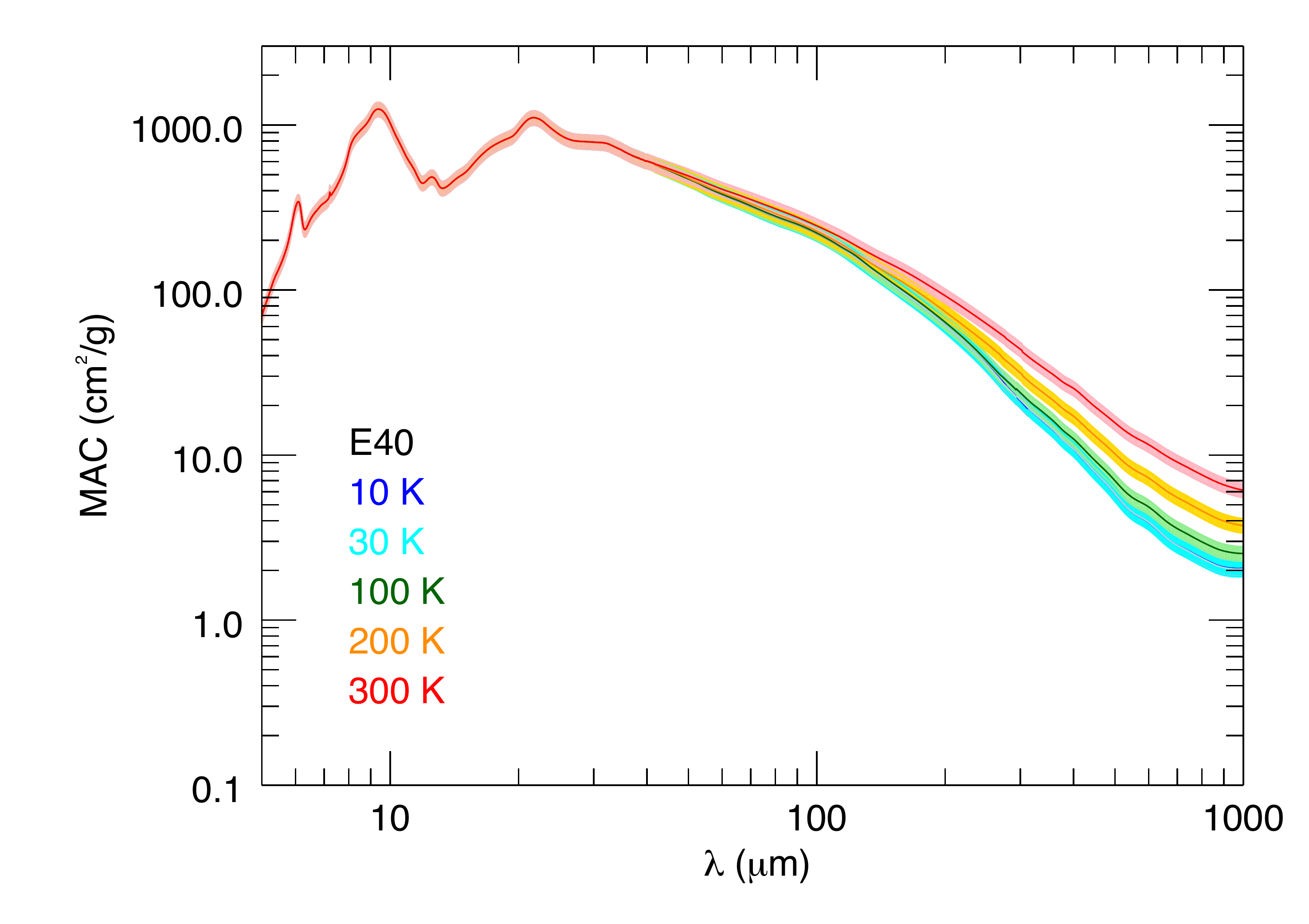}
\includegraphics[scale=.29,trim={0cm 1cm 0cm 1cm},clip]{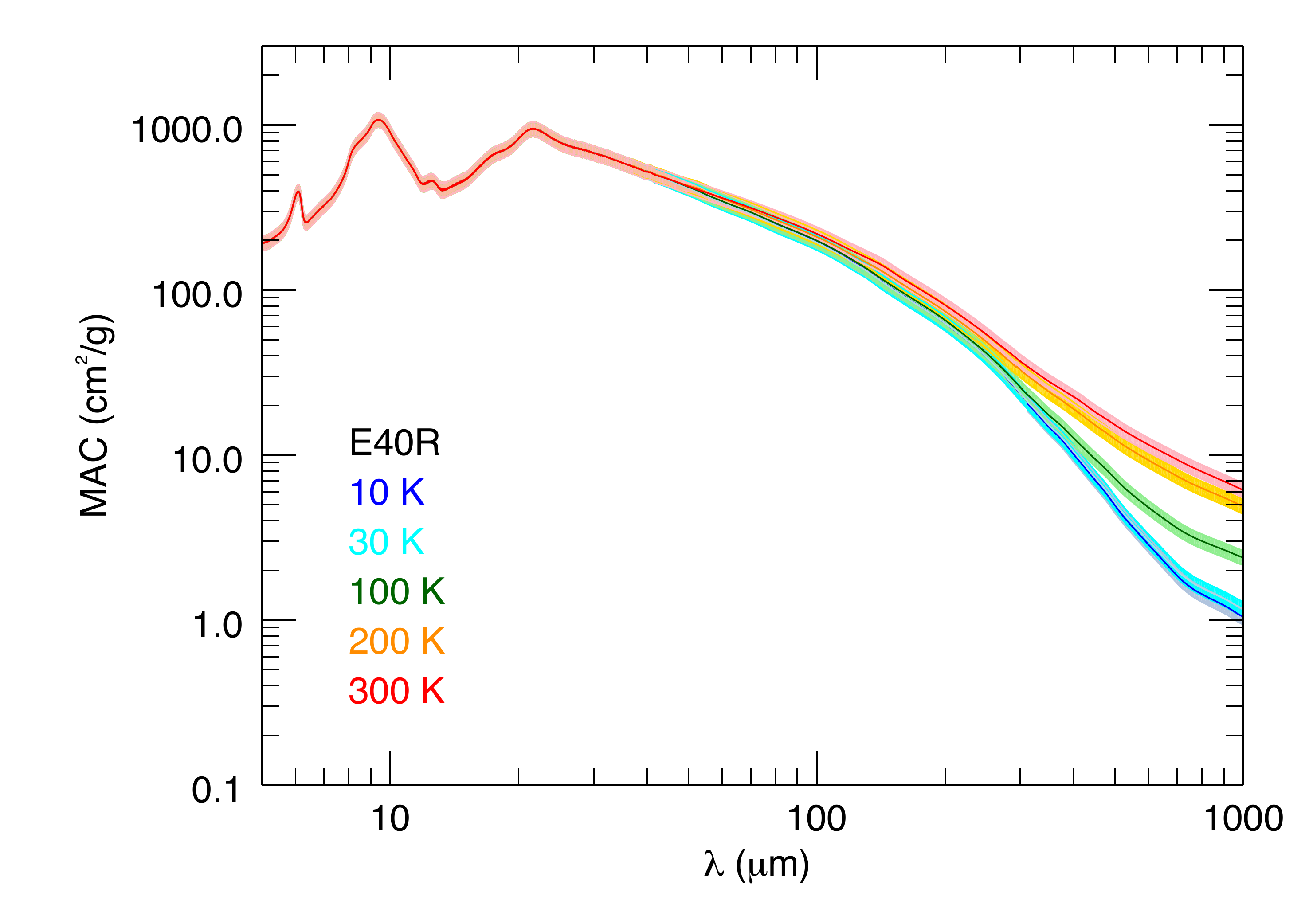}
 \end{center}
         \caption{MAC of the Exx and ExxR samples in the 5 - 1000 $\mu$m domain at room temperature (red curve), 200 K (orange), 100 K (yellow), 30 K (light green) and 10 K (blue). The unprocessed samples, Exx, are shown in the left panels (from top to bottom: E10, E20, E30 and E40) and the processed samples, ExxR, in the right panels (from top to bottom: E10R, E20R, E30R and E40R). The shaded area represents the uncertainty on the experimental data. }
             \label{kappafir}
\end{figure*}

The MAC of all the samples, measured from room temperature down to 10 K (in the range 30 - 1000 $\mu$m), are shown in Fig.~\ref{kappafir} in the 5 - 1000 $\mu$m spectral domain. In the FIR domain, the MAC of the eight samples exhibits the same dependency with temperature: Above 30~K the MAC increases with the grain temperature. We do not observe any change of the MAC from 10 K to 30 K for any of the samples. The wavelength above which the variation of the MAC with temperature is detectable depends on the sample and on the grain temperature. For a given sample, it appears at a shorter wavelength at high temperature than at low temperature. Typically, the MAC variations with temperature are visible for wavelengths longer than $\sim$ 100/200 $\mu$m depending on the samples studied. 

\begin{figure*}[!th]
\begin{center}
\includegraphics[scale=.19, trim={0cm 0.5cm 0cm 0cm}, clip=true]{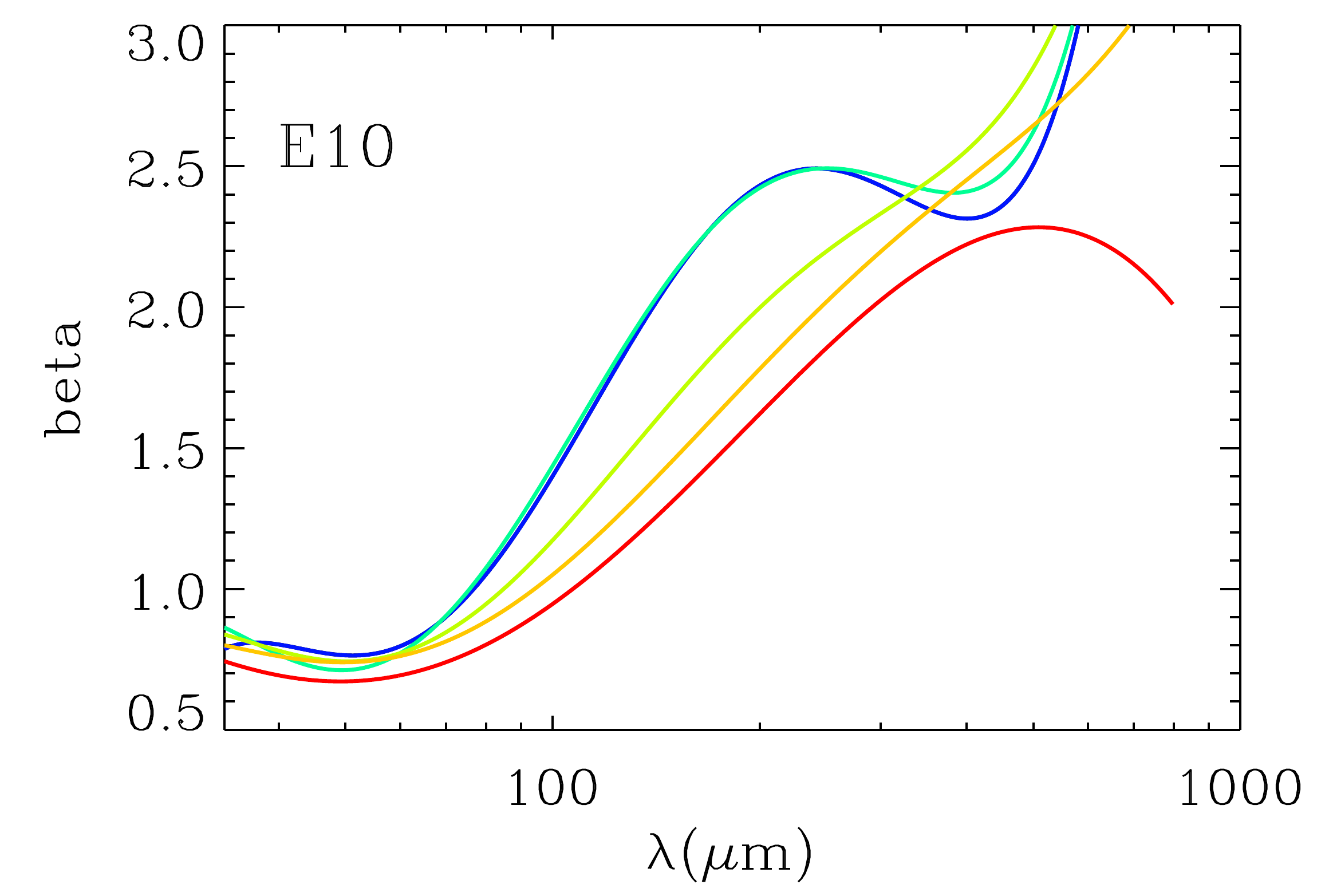}
\includegraphics[scale=.19, trim={0cm 0.5cm 0cm 0cm}, clip=true]{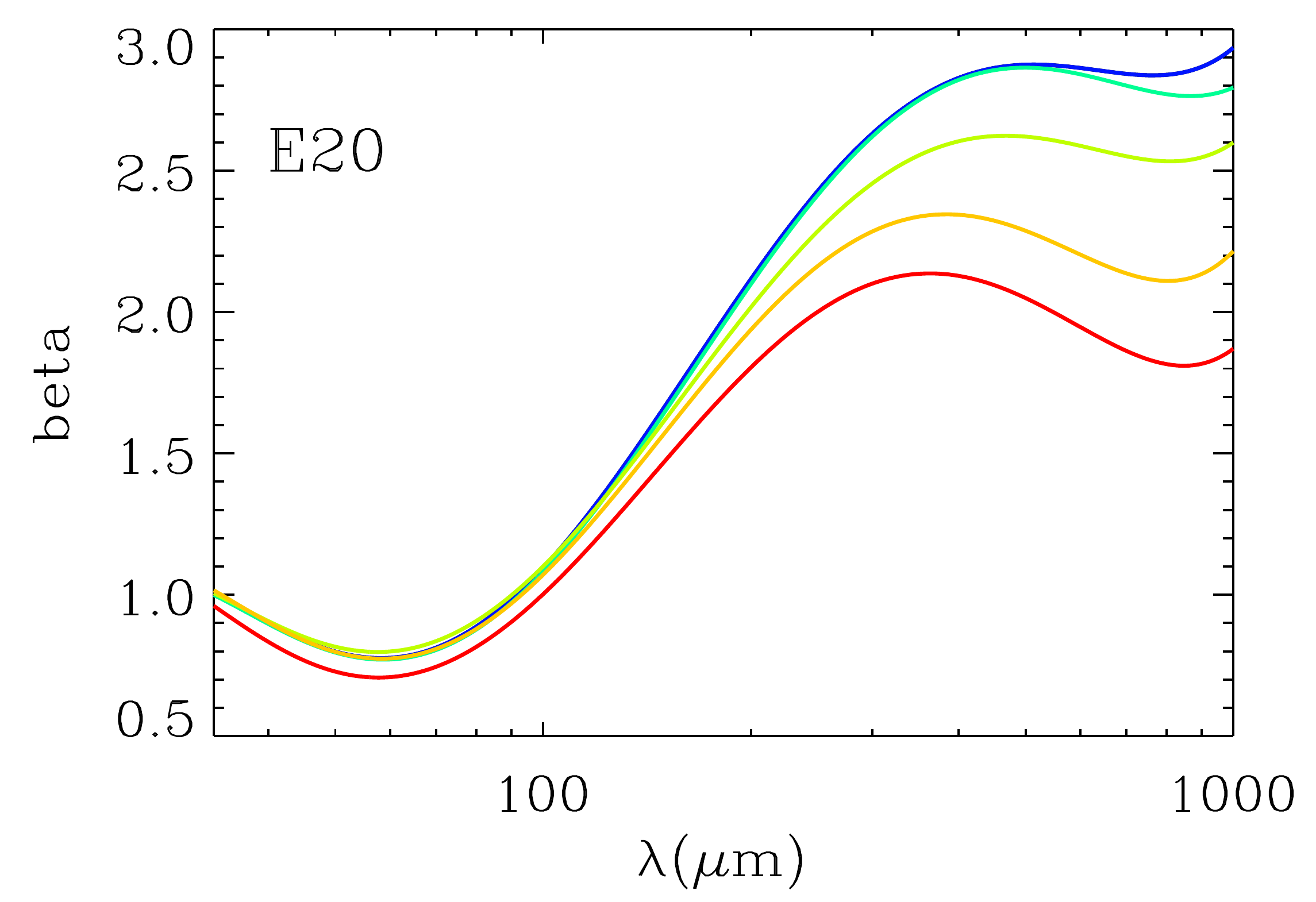}
\includegraphics[scale=.19, trim={0cm 0.5cm 0cm 0cm}, clip=true]{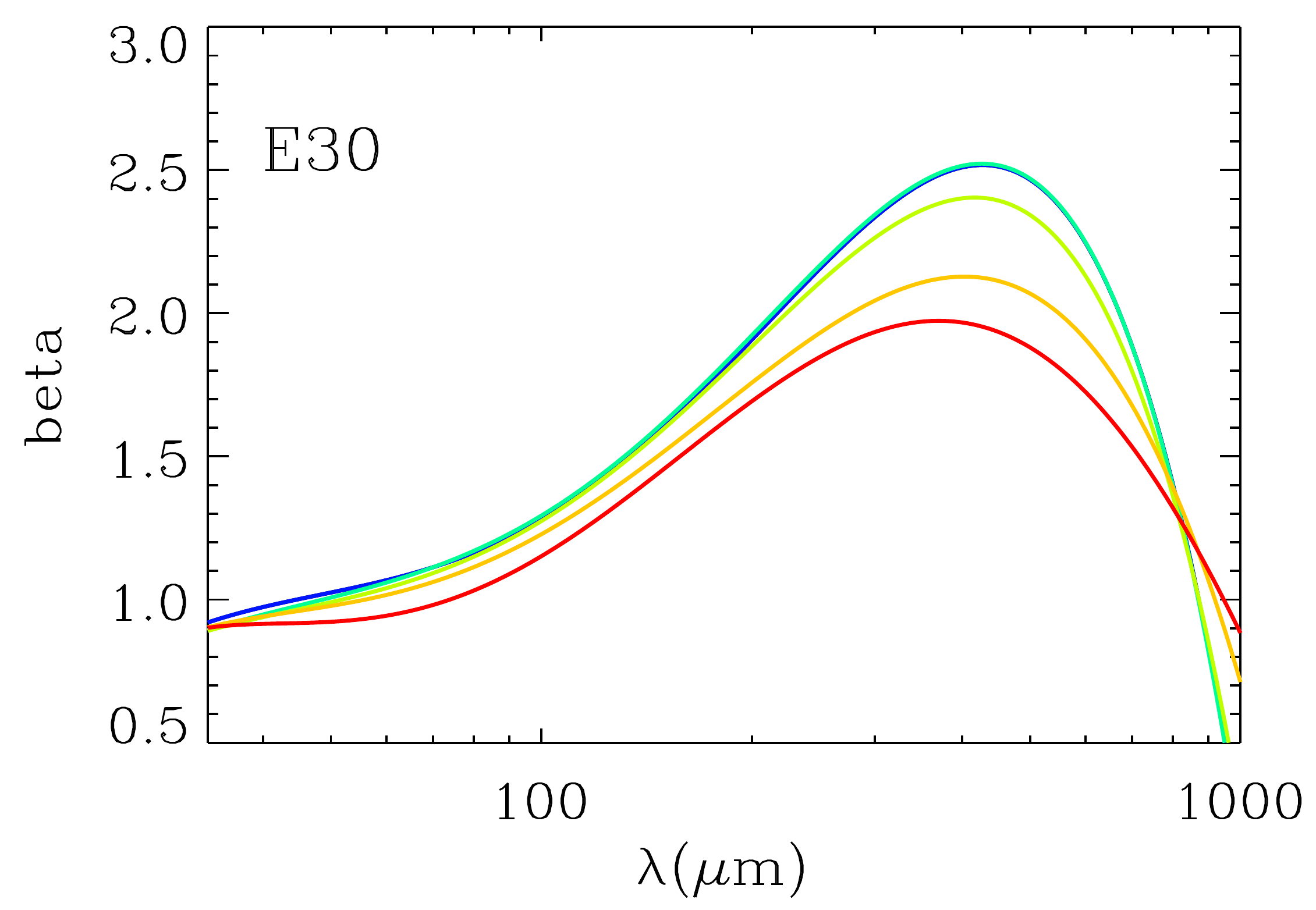}
\includegraphics[scale=.19, trim={0cm 0.5cm 0cm 0cm}, clip=true]{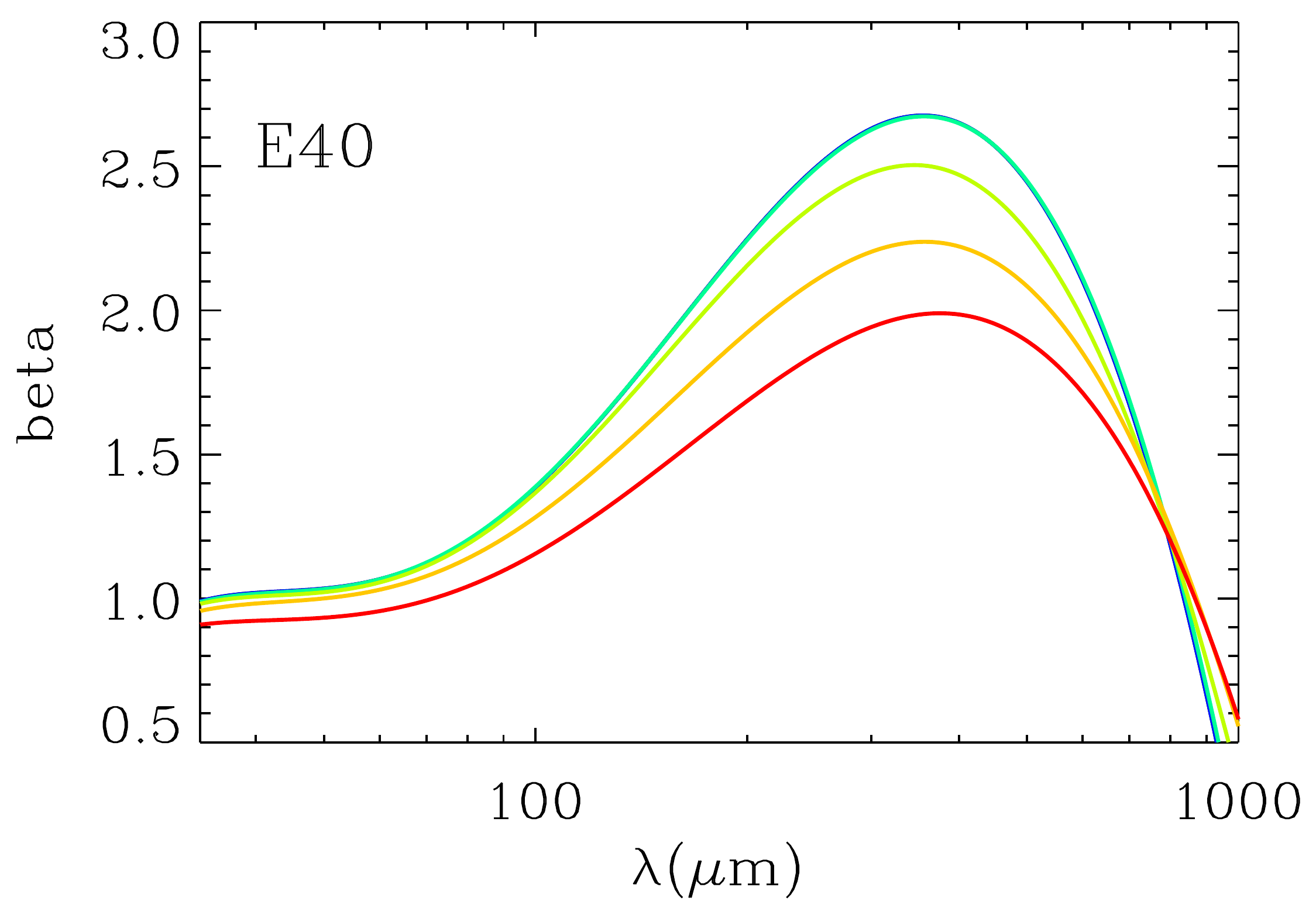} \\
\includegraphics[scale=.19, trim={0cm 0.5cm 0cm 0cm}, clip=true]{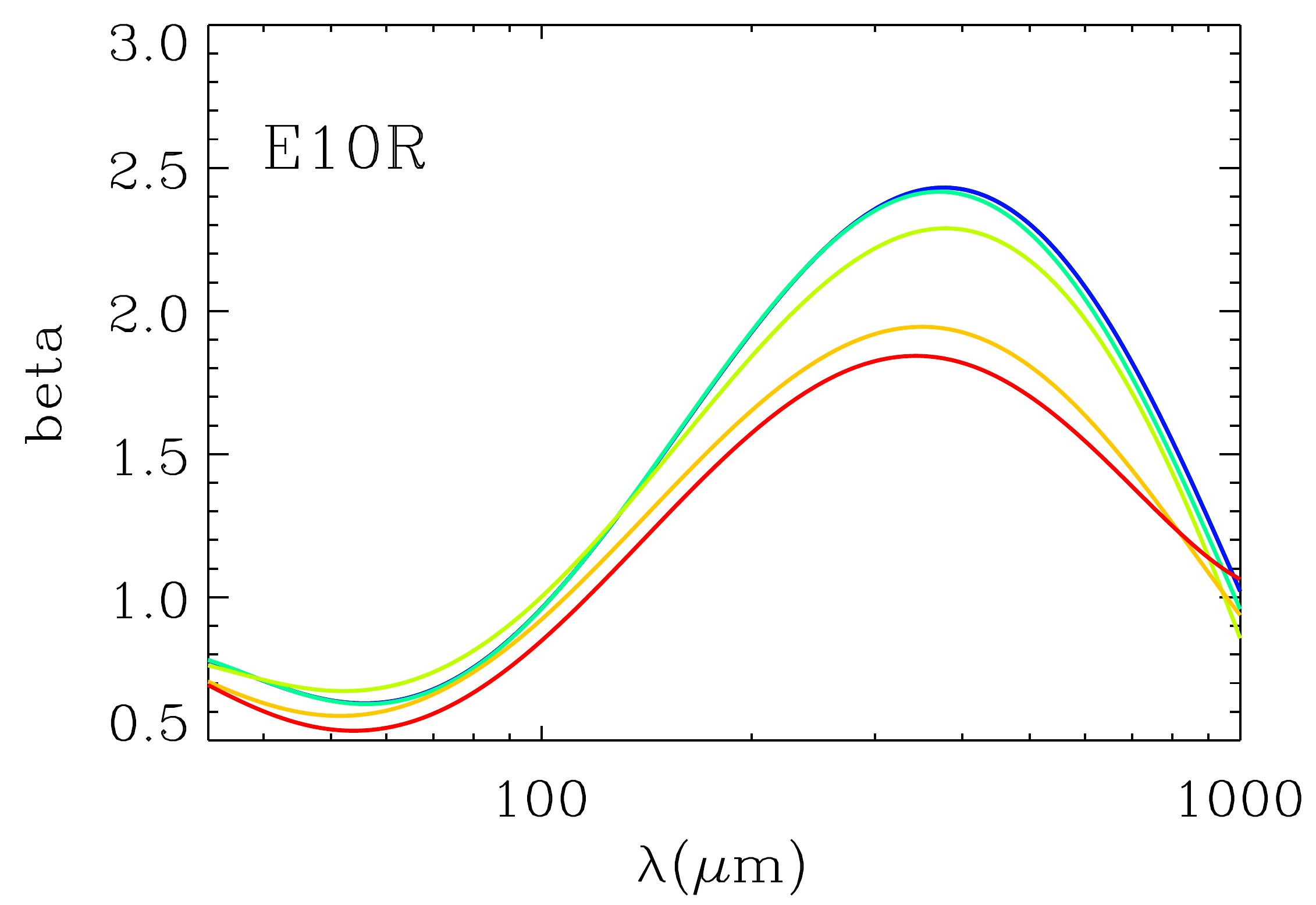}
\includegraphics[scale=.19, trim={0cm 0.5cm 0cm 0cm}, clip=true]{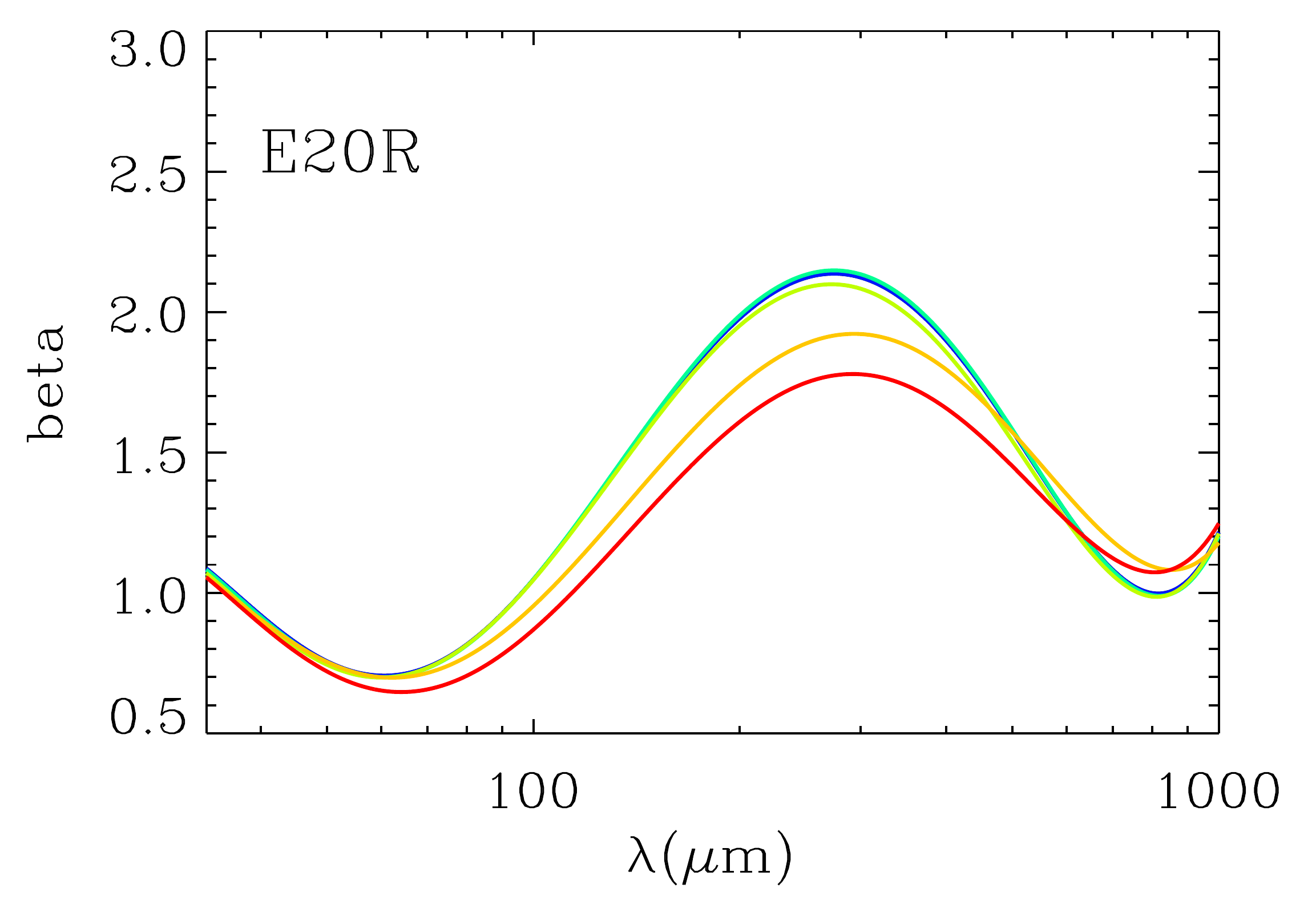}
\includegraphics[scale=.19, trim={0cm 0.5cm 0cm 0cm}, clip=true]{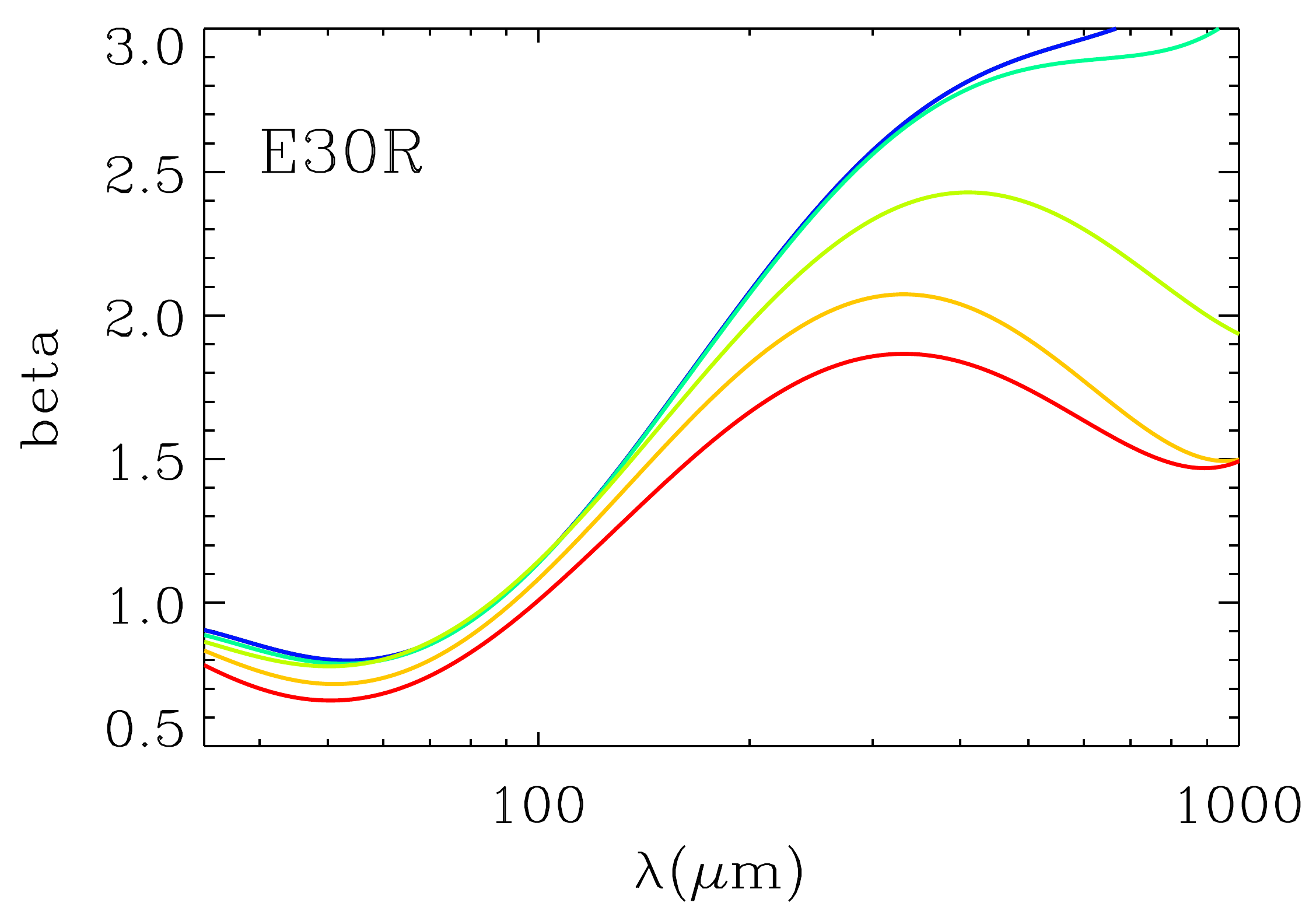}
\includegraphics[scale=.19, trim={0cm 0.5cm 0cm 0cm}, clip=true]{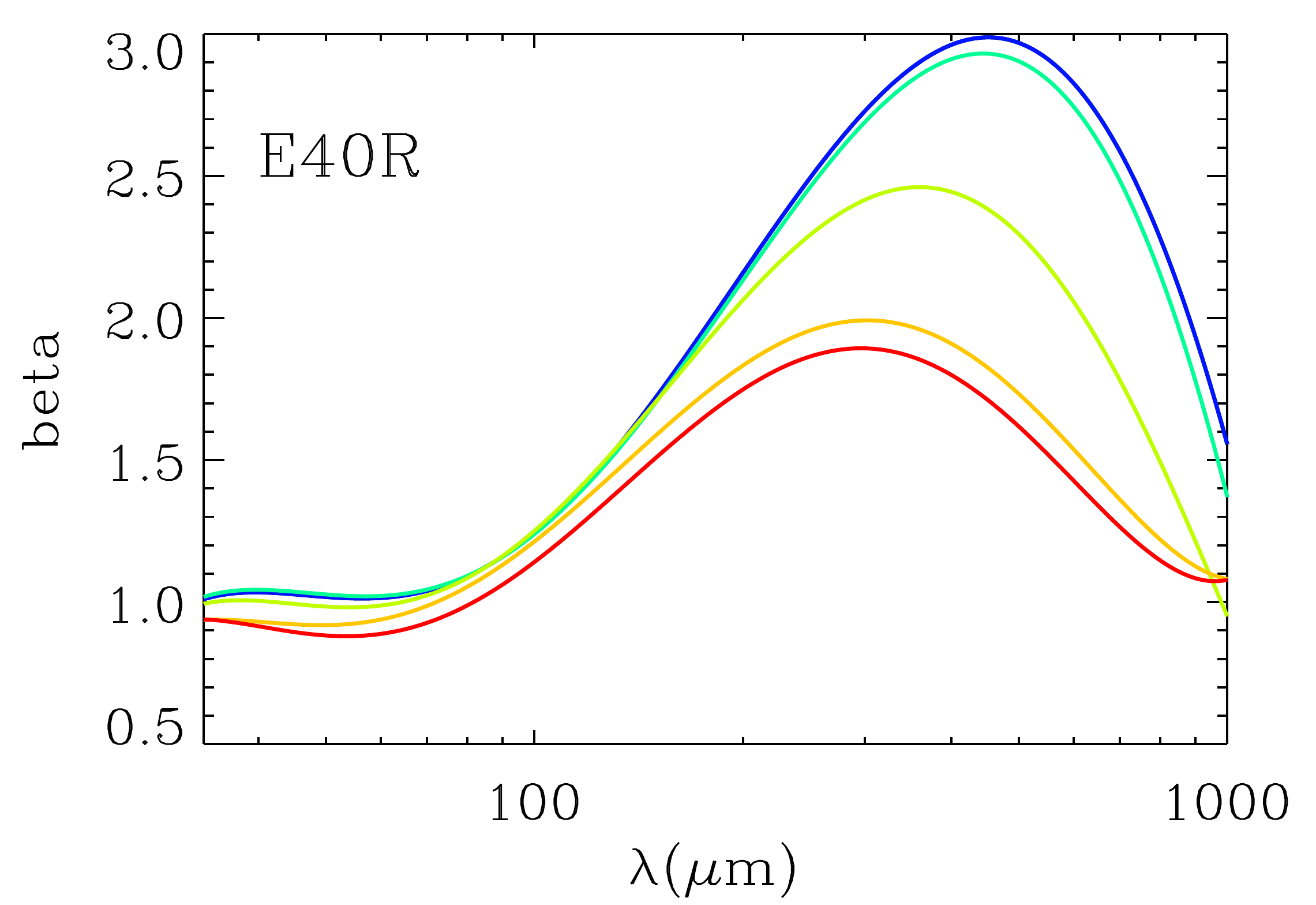}
 \end{center}
         \caption{Beta value calculated at each wavelength from the fit of the MAC curves (see Sect.~\ref{spectro}) for the measured temperatures: 300 K (red), 200 K (orange), 100 K (green), 30 K (light blue) and 10 K (blue).}
            \label{beta}
\end{figure*}

This variation of the MAC intensity is accompanied by a change of its spectral shape which cannot be reproduced by a single power law in ${\lambda}^{-\beta}$, where the spectral index, $\beta$, is defined as the slope of the MAC in the FIR/submm in the log-log representation. To derive $\beta$ at each wavelength, we have fit the MAC of each sample, at each temperature, with a sixth order polynomial in the wavelength range 30 - 1000 $\mu$m and calculated the derivative (Fig.~\ref{beta}). This emphasizes that, at a given temperature, $\beta$ is changing with wavelength. In addition, at a given wavelength in the FIR/submm, $\beta$ increases when the grain temperature decreases, and the amplitude of the variation of $\beta$ with temperature is higher at long wavelengths. In the red wing of the bending vibrational band, around $\sim$ 30 - 80 $\mu$m, the value of $\beta$ decreases below 1 whereas, above $\sim$ 100 $\mu$m, it increases up to values greater than 2. This reflects the presence of the "bump" observed in the MAC spectra with a peak maximum around 100 $\mu$m and is further discussed in Sect.~\ref{comp_labo}. The values of the MAC of all samples are reported in Table~\ref{table_kappa} for a selection of wavelengths. At 100 $\mu$m, the MAC varies little with temperature, no more than 10  to 17 \% from 10 K to 300 K depending on the samples. As the wavelength increases, the variation of the MAC gets stronger and, at 1 mm, it varies by a factor in the range 1.7 to 6 from 10 K to 300 K according to the samples. \\

\begin{figure*}[!h]
\begin{center}
\includegraphics[scale=.3,trim={0cm 1cm 0cm 1cm},clip]{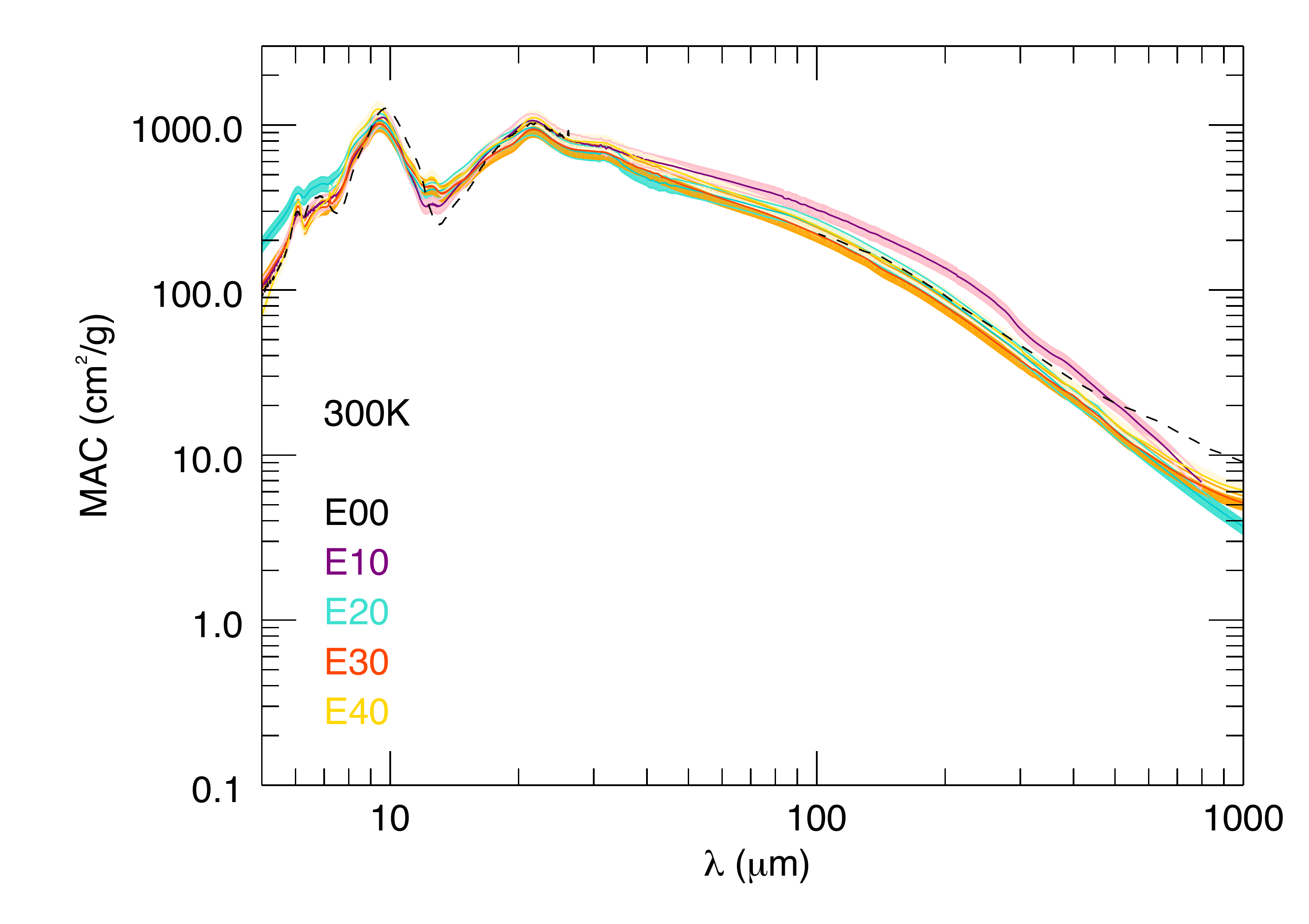}
\includegraphics[scale=.3,trim={0cm 1cm 0cm 1cm},clip]{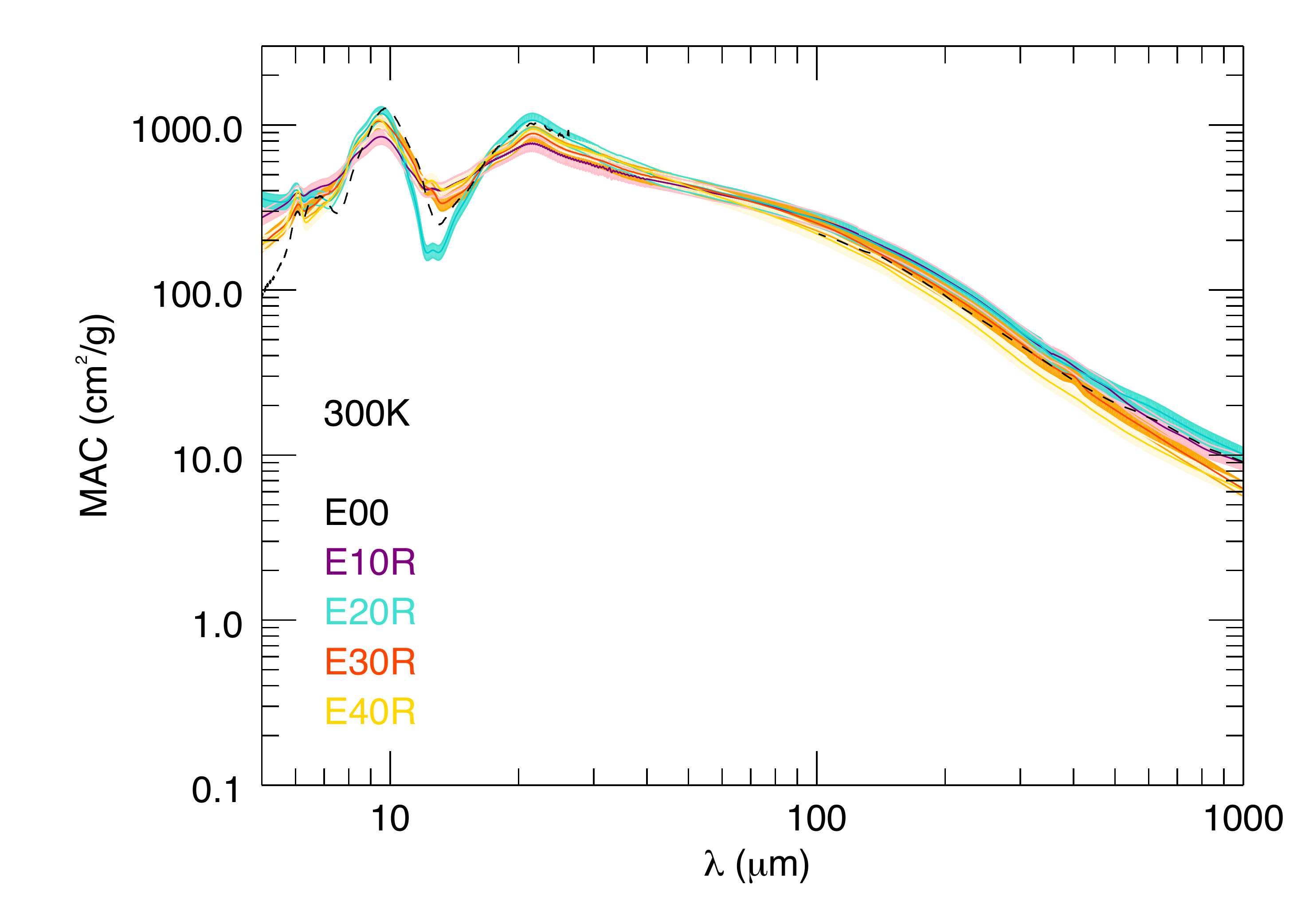}\\
\includegraphics[scale=.3,trim={0cm 1cm 0cm 1cm},clip]{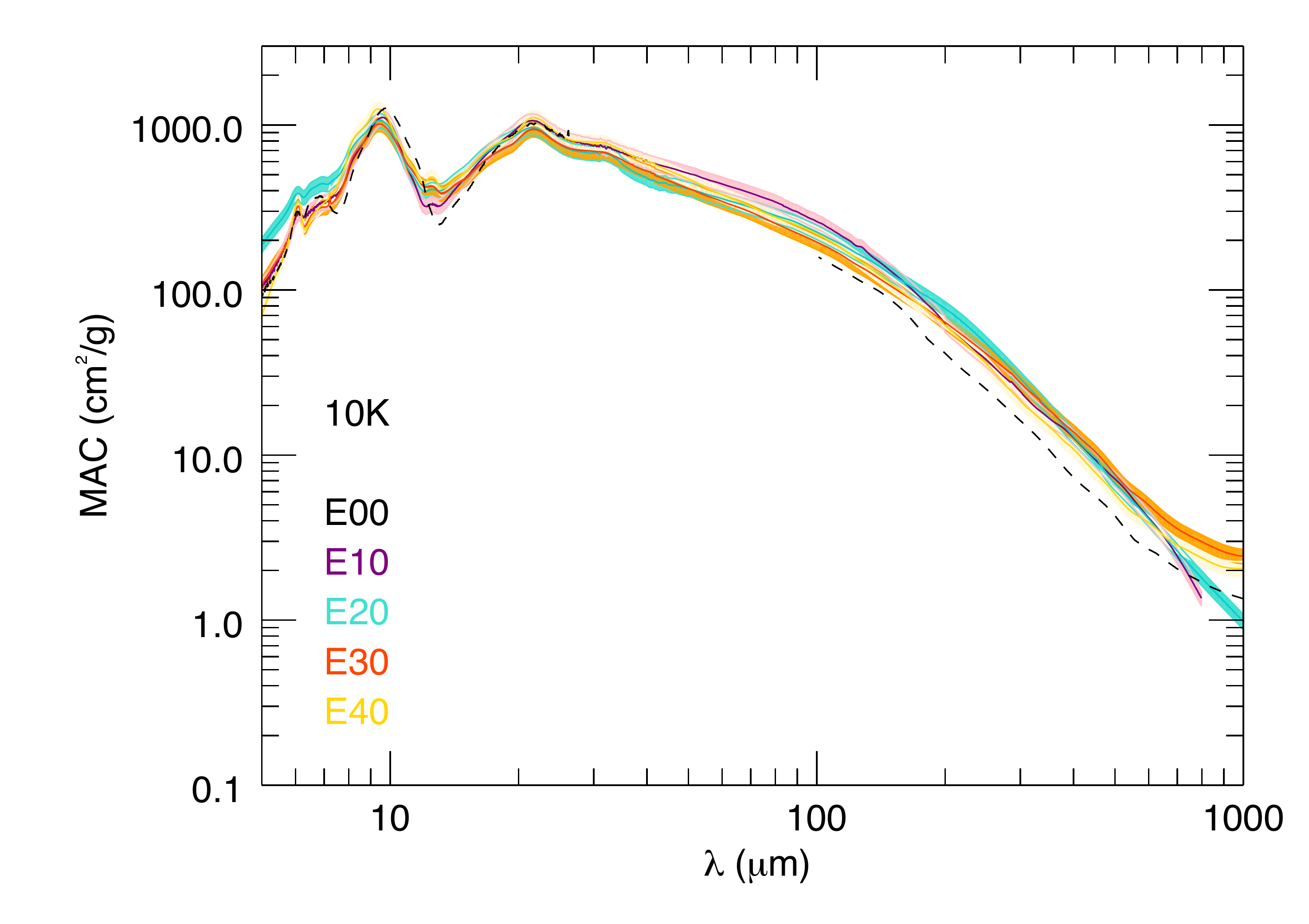}
\includegraphics[scale=.3,trim={0cm 1cm 0cm 1cm},clip]{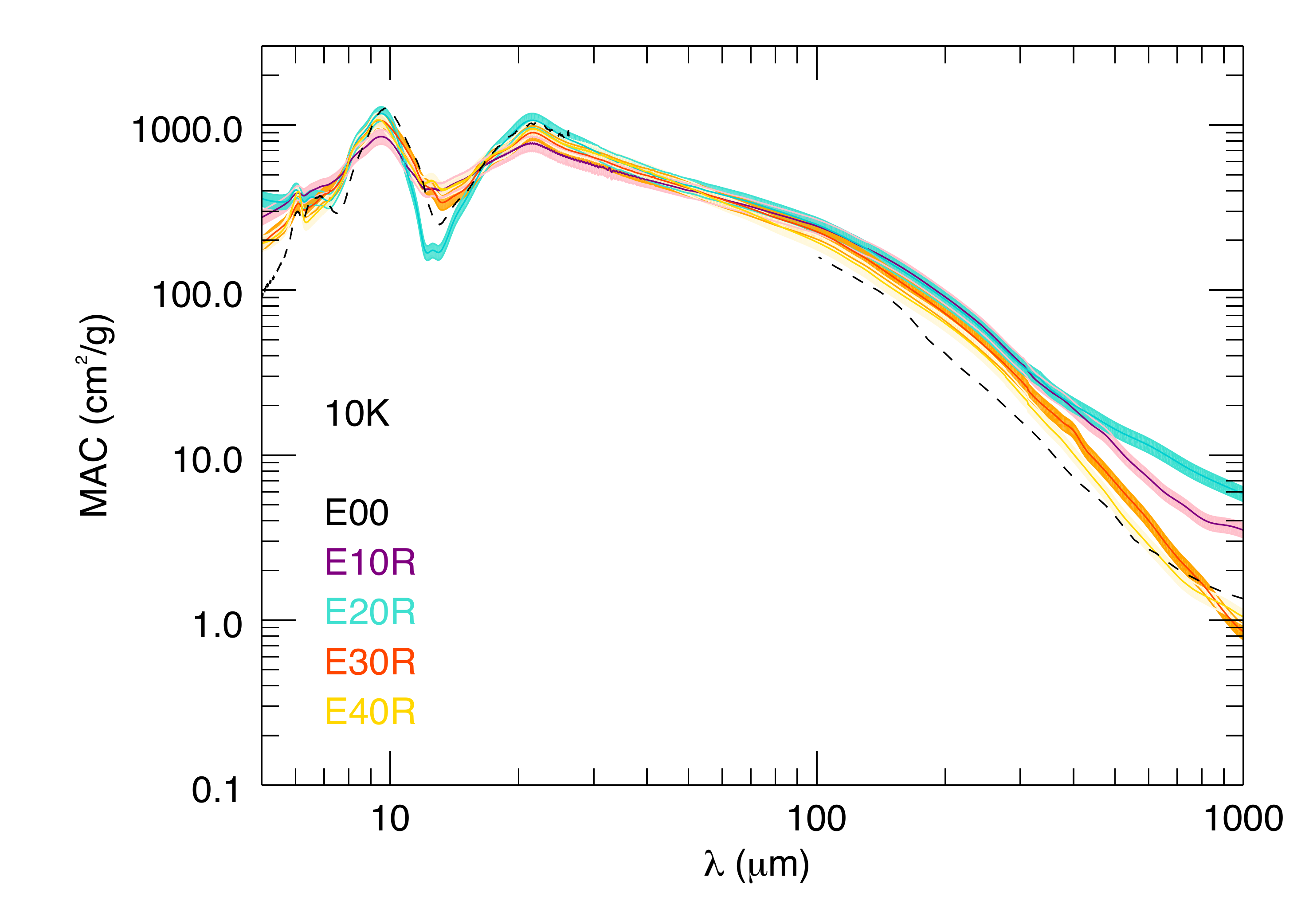}\\
 \end{center}
         \caption{Comparison of the MAC of the samples E10, E20, E30 and E40 (left panels) and of the MAC of the E10R, E20R, E30R and E40R samples (right panels) at 300 K (top panels) and at 10 K (bottom panels). The MAC of the E00 sample is shown as the dotted black line.}
             \label{kappa_TbyT}
\end{figure*}

\begin{figure}[!t]
\begin{center}
\includegraphics[scale=.35, trim={0.5cm 2.5cm 0.5cm 2cm},clip]{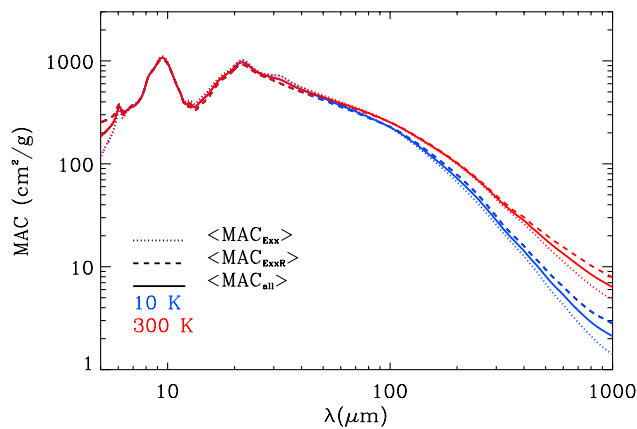}
 \end{center}
         \caption{Average MAC of the samples at room temperature (red curve) and at 10 K (blue). The MAC is averaged over the four Exx samples, < MAC$\mathrm{_{Exx}}$> (dotted lines); over the four ExxR samples, < MAC$\mathrm{_{ExxR}}$> (dashed lines) and over all the samples, < MAC$\mathrm{_{all}}$>  (continuous lines). }
             \label{kappa_mean}
\end{figure}

\begin{figure*}[!h]
\begin{center}
\includegraphics[scale=.26, trim={3cm 2.5cm 2cm 1cm},clip]{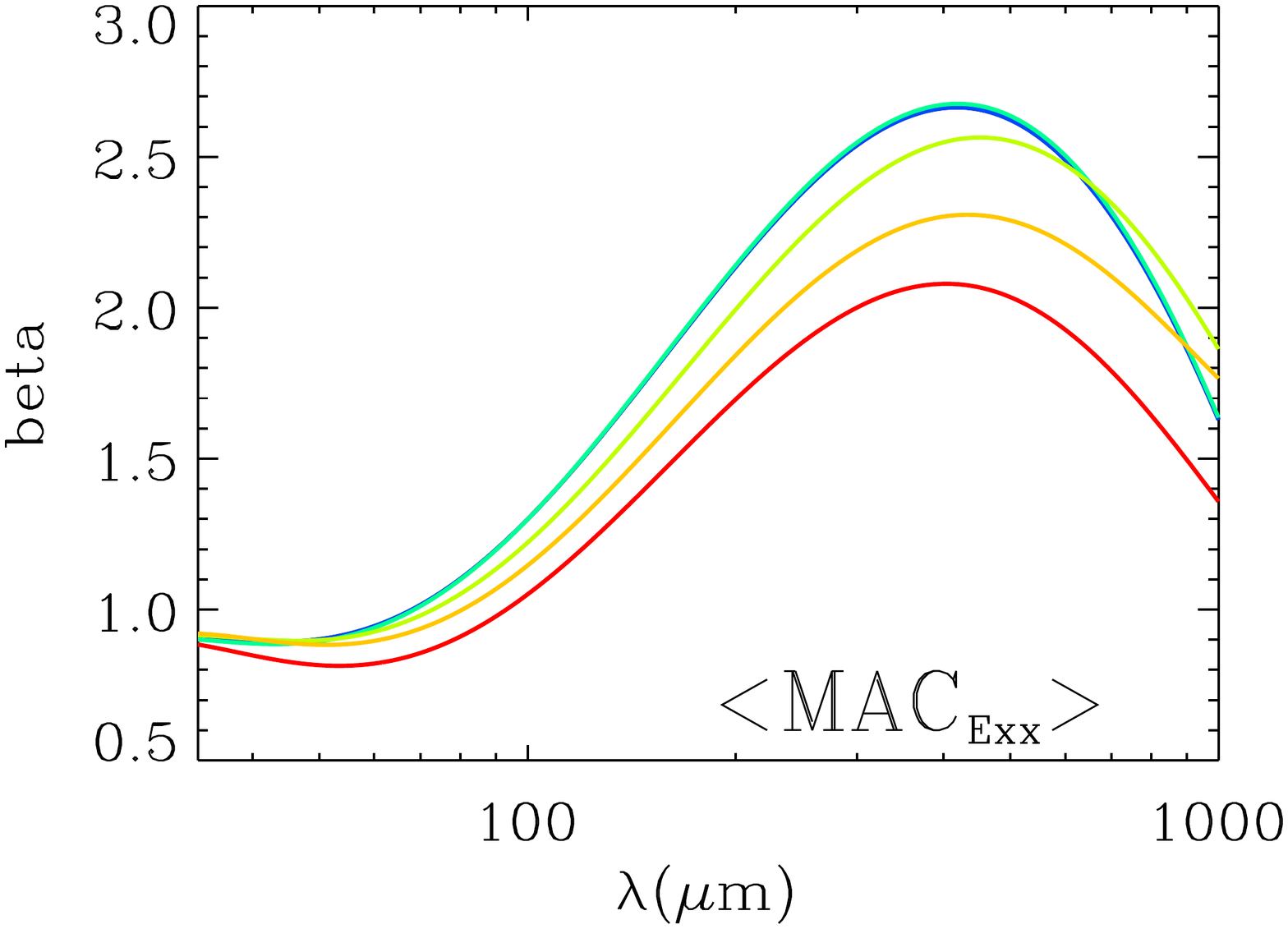}
\includegraphics[scale=.26, trim={3cm 2.5cm 2cm 1cm},clip]{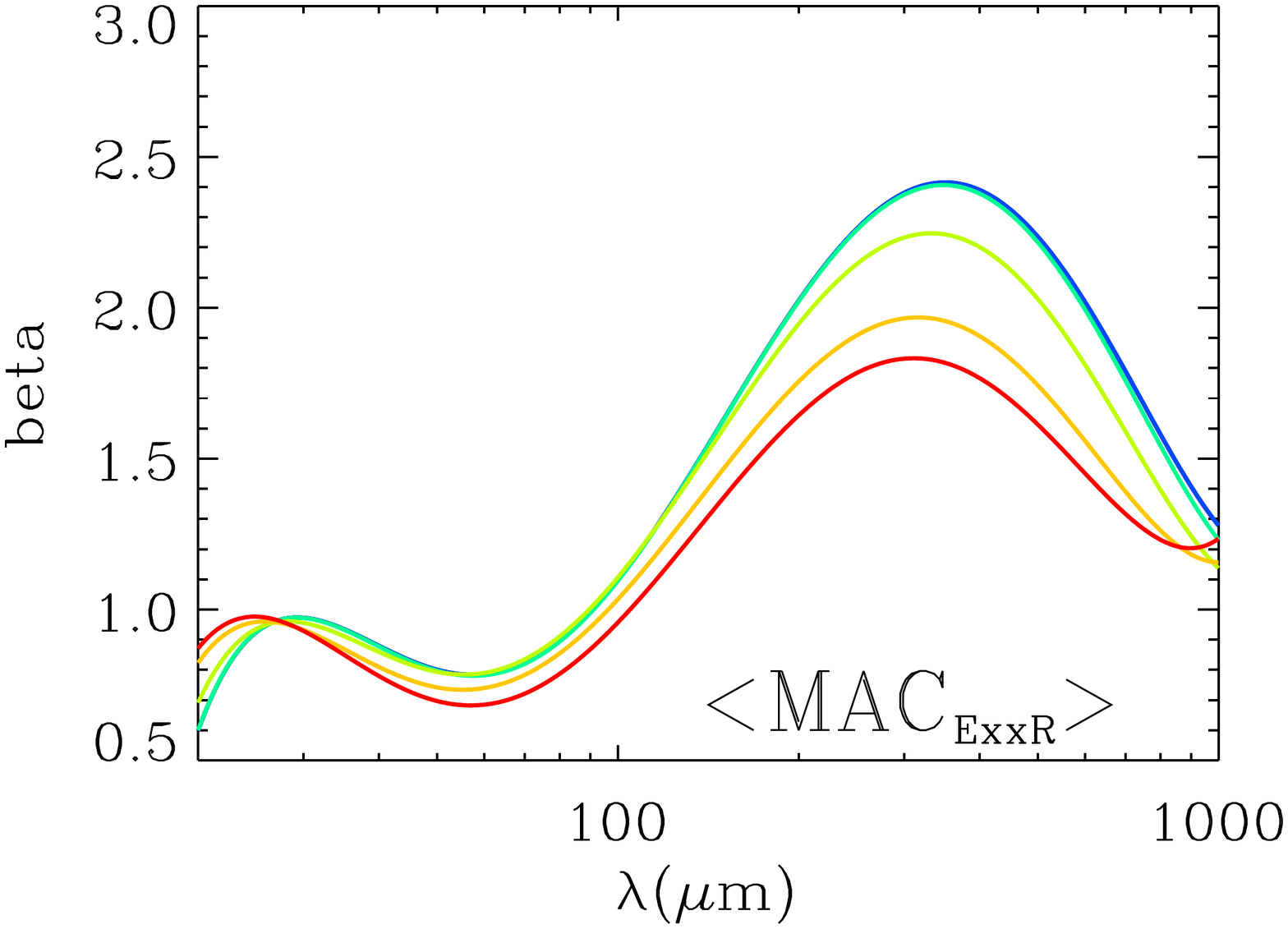}
\includegraphics[scale=.26, trim={3cm 2.5cm 2cm 1cm},clip]{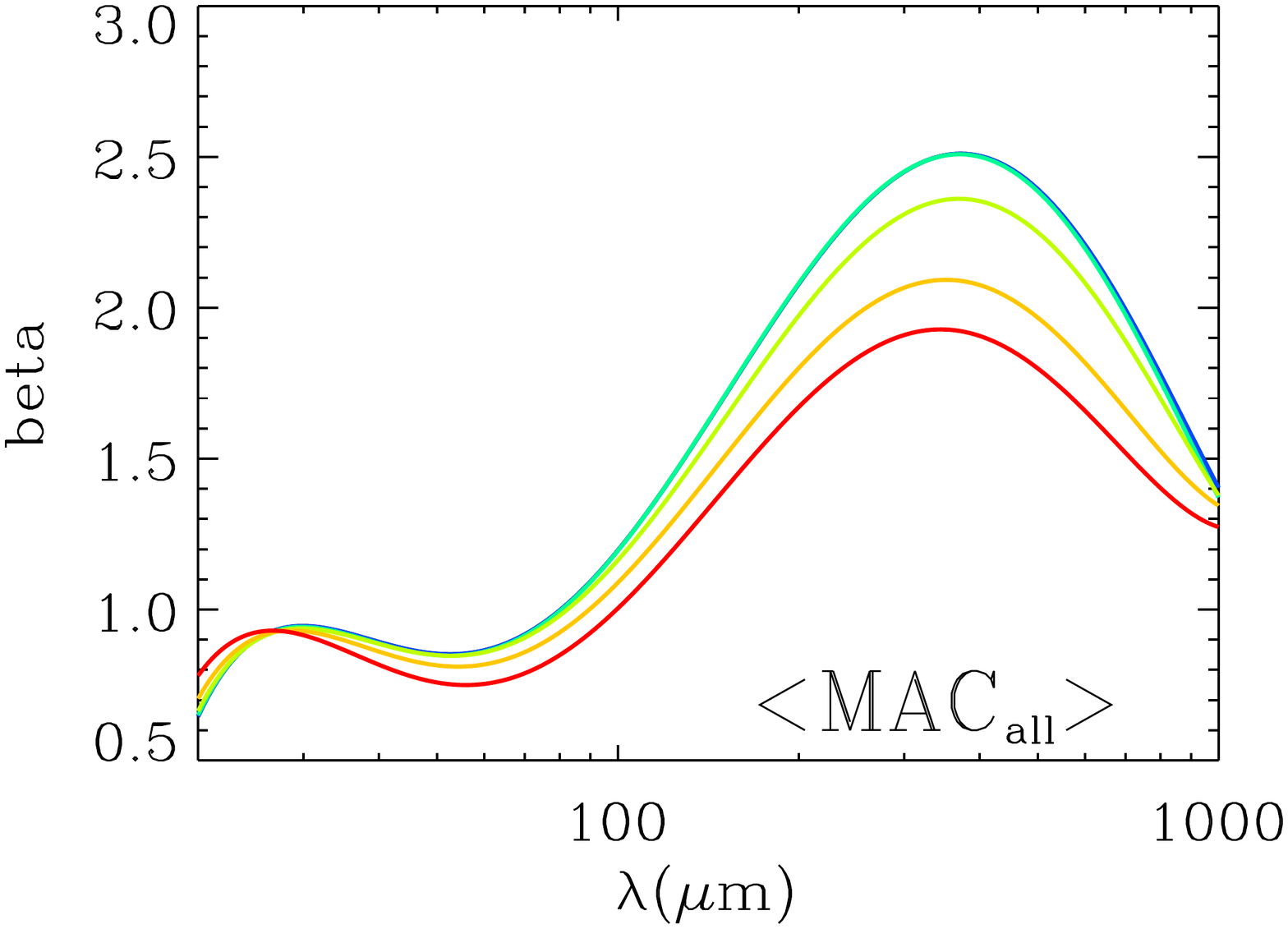}
 \end{center}
         \caption{Beta value calculated at each wavelength from the fit of the MAC curves averaged over the E10, E20, E30 and E40 samples (left panel), the  E10R, E20R, E30R and E40R samples (middle panel) and all the samples (right panel) at the measured temperatures: 300 K (red), 200 K (orange), 100 K (green), 30 K (light blue) and 10 K (blue).}
            \label{beta_mean}
 \end{figure*}

\section{Discussion}
\label{discussion}

\subsection{Influence of the composition and processing of the analogues on the MAC}
\label{MAC_proc_comp}

Figure~\ref{kappa_TbyT} compares, for a given temperature (10 K and 300 K), the spectra of the samples with different iron contents, for each series of samples, Exx and ExxR. The MAC of the pure Mg-rich sample, E00, is also added for comparison. The MAC of the unprocessed samples E00, E10, E20, E30 and E40, look remarkably similar below $\sim$ 700/800 $\mu$m whereas above $\sim$ 700/800 $\mu$m, the MAC of the four samples are  different. The MAC of samples E10 and E20 gets steeper at long wavelengths whereas the MAC of samples E30 and E40 flattens. The MAC of the E00 sample is closer to samples E30 and E40 than to samples E10 and E20. As for the unprocessed samples, the MACs of  E10R, E20R, E30R and E40R are very similar at short wavelengths whereas above 300/400 $\mu$m, depending on the temperature, the MACs of the four processed samples differ. The MAC of samples E10R and E20R flattens at long wavelengths whereas the MAC of samples E30R and E40R steepens. Hence, no clear trend emerges from these measurements about the influence of the iron content on the MAC of the samples, for either of the two series. However, we know from the results of the MIR and M\"ossbauer spectroscopic measurements that the samples of each series contain a small amount of iron oxides (less than 5 - 10\%, see Sect.~\ref{carac}). This certainly complicates the interpretation of the FIR spectroscopic data, since in this domain, the MAC depends on the structure at microscopic scale of the materials, a structure which is influenced by the form in which the iron is present in the samples (within the silicate network or iron in oxides). 

In the FIR, the difference between the MAC of the samples having various iron content is more important for the processed samples than for the unprocessed samples. This should reflect the effect of the processing applied to the ExxR samples. This processing induces changes of the chemical and structural nature of the samples as indicated by the M\"ossbauer and MIR spectroscopic results (Sects.~\ref{carac} and ~\ref{spectro}). However, the modifications of the MAC induced by the processing do not show a common behavior in terms of spectral shape and intensity. It is different from one sample to another. For example, the MAC of the unprocessed samples E10 and E20 are weaker than the MAC of the processed samples E10R and E20R whereas this is the opposite for samples E30/E30R and E40/E40R. The similarity of the MAC of the samples containing 10 \% and 20 \%  iron on one side, and  30 \% and 40 \% iron on the other side, both for the normal and processed series, is probably related to the fact that, as indicated in Sect.~\ref{carac}, the samples E10 and E20 have comparable composition, as do samples E30 and E40. In addition, the two pairs of samples E10/E20 and E30/E40 were synthesized in separate runs. Consequently, the samples of each pair should be very similar in terms of structure or chemical and/or structural homogeneity at microscopic scale, whereas it is possible that the two pairs differ slightly. 

Despite the fact that the processing of the samples does not show any clear trend, it has a non negligible effect, which is emphasized when comparing the MAC averaged over the four processed samples,  <MAC$_\mathrm{ExxR}$>,  with the MAC averaged over the four unprocessed samples, <MAC$_\mathrm{Exx}$>, and the MAC averaged over the eight samples <MAC$_{\mathrm{all}}$>  (Fig.~\ref{kappa_mean} and Table~\ref{table_kappa}). In addition it is clear from Fig. \ref{kappa_mean} that averaging the MAC over different combinations of samples does not suppress the temperature variation of the MAC in the FIR domain. In addition, even though the average MAC is smoother than the MAC of each sample, they retain a complex spectral shape incompatible with a single power law  (Fig.~\ref{beta_mean}).

\subsection{Comparison with previous experimental data}
\label{comp_labo}

The observed behavior of the MAC with temperature and its complex spectral shape are similar to those already observed for the MAC of amorphous silicate materials by \citet{mennella1998, boudet2005,coupeaud2011,demyk2017}. All these studies show that the MAC of amorphous silicate analogues is temperature dependent, that it increases when the grain temperature increases, and that the spectral shape of the MAC cannot be fitted with a single power law in ${\lambda}^{-\beta}$. Most of the samples from these studies are iron free and the overall qualitative agreement of all these samples thus reflects that, more than the composition (Mg vs. Fe content), the structure at nanometer and atomic scales (e.g., degree of polimerization of the Si tetrahedra, presence of defects, presence of Si atoms not tetrahedrally linked to oxygen in a chaotic structure) governs the FIR/submm absorption. 

\cite{mennella1998} present a study of an Fe-rich amorphous silicate at low temperature in the FIR/submm range (20 $\mu$m - 2 mm). They synthesized  amorphous Fe-rich olivine-like grains by laser vaporization of the natural crystalline target mineral (fayalite), in an oxygen atmosphere, ending up with a sample of composition Mg$_{0.18}$Fe$_{1.82}$SiO$_4$ in the form of small spherical particles of diameters in the range 13 - 35 nm, aggregated in chains for the smallest particles. The MAC of this sample, named FAYA, is similar with our measurements. The MAC values are in the same range as the Exx and ExxR samples in the MIR and in the FIR. At 1mm, the MAC of the FAYA sample is $\sim$ 5.0 cm$^2$.g$^{-1}$ and $\sim$ 0.86 cm$^2$.g$^{-1}$ at 300 K and 24 K respectively. The main difference between this sample and the analogues studied here lies in the shape of the MAC curves. The absence of  a band at $\sim$ 33 $\mu$m in the FAYA spectrum suggests that the iron contained in the FAYA sample is not in the same iron oxide phase as in the Exx samples. The prominent bump in the range 100 - 300 $\mu$m, present in our samples, is not observed in Mennella's study. Such a bump was also observed in the Mg-rich glassy silicates from \citet{demyk2017} and explained in terms of Boson peak (BP).  Such BP appears to be a universal feature of solid state of different amorphous materials, nevertheless without any clear and widely shared understanding of the underlying physical process. The analytical model from \cite{gurevich2003} describes the BP as an excess in the vibrational density of states, $g_{BP}(\omega)$, over the usual Debye vibrational density of states, $g_{Debye}(\omega) \propto\omega^2$. Although it gives good fits of the MAC of the Mg-rich glassy silicates from \citet{demyk2017} in the range 30 - 700 $\mu$m, it fails to reproduce the MAC of the ferromagnesium amorphous samples Exx and ExxR. The capacity of this model to produce a reasonable fit is restored if the Gurevich's BP profile overlaps with an absorption process $ \propto\omega^n$, with n in the range 0 to 2, depending of the sample, but with no clear correlation with any sample characteristics. Clearly these experimental MAC curves fail to be modeled with any current physical model, and should be taken as experimental data to be considered by theoretical solid-state physicists.

\cite{richey2013} measured the transmission and reflexion spectra of FeSiO silicates produced in a condensation flow apparatus at low temperature (5 - 300 K) in the 15.4 - 330 $\mu$m  domain. Dust produced in such an apparatus consists in small nanograins (2 - 30nm) aggregated in the form of fluffy, open clusters containing hundreds to thousands of grains. From these spectra \citet{richey2013} calculated the optical constants of the FeSiO sample and they discussed their results in terms of n and k, the refractive index and the absorption coefficient, respectively, rather than in terms of MAC as in this study. They observe a small temperature variation in the absorption coefficient, k, in the vibrational band at 21.5  $\mu$m and a variation of $\sim$ 5\% in k, above 100 $\mu$m, in the temperature range 100 - 300 K. In the 30 - 100 $\mu$m, these results on the absorption coefficient of the FeSiO sample are compatible with our results observed on the MAC of  sample Exx and ExxR which show little or no variation.

These various measurements show that the MAC dependence on temperature appears above 100 $\mu$m with only very small variation in the MIR  domain. This is true whatever the structure of the material, from the most highly disordered silicates from \cite{richey2013} to the disordered, but less "chaotic" silicates from \cite{mennella1998} and the present study.

\subsection{Comparison with cosmic dust models and astrophysical implications}
\label{dust_model}

Cosmic dust models are built to interpret astronomical observations either to study dust itself or to use dust as a probe of the astrophysical environment. Most dust models consider two main dust components, the carbonaceous dust and the silicate dust, which may themselves be divided up into several dust populations. The carbonaceous dust component usually includes the very small grains (VSG, a $\le$ 15 nm) and the polycyclic aromatic hydrocarbons (PAHs). Some models also include a population of large carbonaceous grains (a $\ge$ 50 nm) composed of graphite \citep{draine1984} or amorphous carbon grains more or less aliphatic \citep{compiegne2011,jones2013}. All cosmic dust models include a population of large grains made of silicates, which are generally pure magnesian silicates. The "astrosilicates" from \citet{li2001} are based on experimental data of Mg-rich silicates in the UV/VIS domain, on astronomical observations of the ISM dust in the MIR, and on the extrapolation of these data in the FIR/submm \citep[see][for the details]{draine1984}. The "Themis" model \citep{jones2013} considers two silicate dust components which are based on experimental data on Mg-rich amorphous silicates of composition MgSiO$_3$ and Mg$_2$SiO$_4$ in the MIR and in the UV/VIS   \citep{scottduley1996}, and which are extrapolated in the FIR/submm. Following the classical description of the amorphous state absorption such as the Debye model \citep{henning1997}, the extrapolation of both models in the FIR/submm assumes that the MAC has an asymptotic behavior in ${\lambda}^{-2}$ for $\lambda$ $\ge$ 20 $\mu$m. To take into account the fact that iron is highly depleted from the gas phase \citep{jenkins2009} the silicate component of the "Themis" model contains inclusions of metallic iron (Fe) and of iron sulfide (FeS) \citep{koehler2014}.

\begin{figure}[!t]
\begin{center}
\includegraphics[scale=.35, trim={1cm 2.5cm 0cm 2cm},clip]{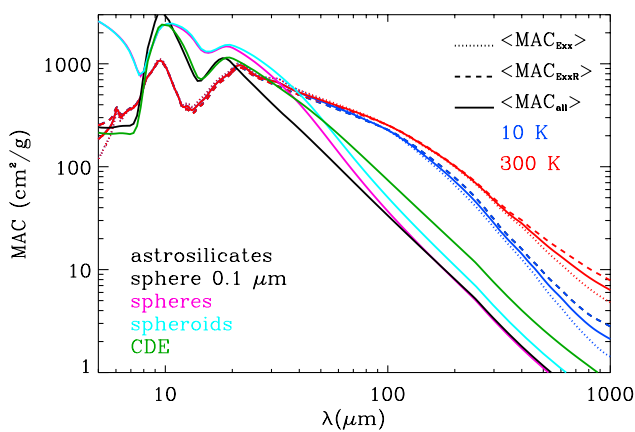}
 \end{center}
         \caption{Comparison of the average MAC at 10 K and 300 K with astronomical dust models. The MAC of the "astrosilicates" from \citep{li2001} are calculated using Mie Theory for a 0.1 $\mu$m size grain (black), for a log-normal grain-size distribution with a mean diameter of 1 $\mu$m for spherical grains (magenta) and for prolate grains (cyan) and for a continuous distribution of ellipsoids (CDE, green). We refer to the text for more details.}
            \label{comp_model_average}
\end{figure}

Figure~\ref{comp_model_average} shows the comparison of the averaged experimental MAC with those calculated from the "astrosilicates" model. The comparison of each sample with the "Themis" dust model and the "astrosilicates" model are shown in Figs.~\ref{comp_model_E10} to ~\ref{comp_model_E40R}.  We have calculated the MAC of a spherical particle of  100 nm in size, the MAC of two populations of particles having a log-normal size distribution centred at 1 $\mu$m (such size distribution should reflect the size of the grains within the pellets), and the MAC of a continuous distribution of ellipsoids (CDE model which assumes that all ellipsoidal shapes are equiprobable, \citet{bohren1998}). The first population of particles consists of spherical grains and the second one of spheroidal grains (prolate grains with an axis ratio of 2). The calculations are performed using Mie theory \citep{bohren1998} for spherical particles and using the DDA code DDSCAT 7.3 developed by \citet{draine2013} for spheroidal particles.

The modeled and measured MACs are very different. In the MIR, where the vibrational bands occur, the discrepancy in the band shape is linked to the composition of dust analogues and to their structure at microscopic scale. As expected, the stretching mode of the Si-O bond of the iron-free samples used in the "Themis" model peaks at longer wavelengths (9.7 $\mu$m and 10.3 $\mu$m for MgSiO$_3$ and Mg$_2$SiO$_4$, respectively) than the iron-rich Exx and ExxR samples (see Table~\ref{MIR}). For the "astrosilicates", which are derived, in the MIR, from astronomical observations, the stretching mode peaks at 9.5 $\mu$m which is close to the peak position of samples E20 and E10R. This suggests that the "astrosilicates" are compatible with silicates containing $\sim$ 10 - 20 \% iron. The bending mode of the Exx and ExxR samples peaks at longer wavelengths than the bending mode of the models reflecting the fact that the structure of the material is different. Related to these structural differences, we note that the Exx and ExxR samples are less absorbant than the dust analogues from cosmic dust models. Interstellar dust is most likely diverse in terms of composition and structure in the various environments in which they are observed. Therefore, these discrepancies do not rule out the studied samples as relevant dust analogues.

Although in the MIR, the dust models are more absorbant than the measured samples, the opposite is true in the FIR. Table~\ref{table_kappa} gives the value of the MAC of the measured samples and of the modeled MAC at selected wavelengths in the 100 $\mu$m - 1 mm range. In this range, the experimental MAC at 300 K is more than five times greater than the modeled MAC for all samples. At 10 K, the experimental MAC is lower than at 300 K and the factor of enhancement compared to the models is smaller, depending on the sample and on the wavelength, however it is always higher than two and usually of the order of four to five. The MAC value of the ferromagnesium silicate analogues in this study is very close the MAC of the pure Mg-rich samples from \citet{demyk2017} and from \citet{coupeaud2011}. This shows that the enhancement of the measured MAC compared to the modeled MAC is not related to the iron content of the grains, that is, to differences in composition between the studied samples and the analogues used in the cosmic dust models. As discussed in detail by \citet{demyk2017}, an enhancement factor greater than two cannot be explained by the effect of grain size, grain shape, or by grain coagulation within the pellets. Grain size and grain shape effects are illustrated in Fig.~\ref{comp_model_average} and the increase of the MAC due to large (micronic) spherical grains is negligible in the FIR. Coagulation might happen during the process of fabrication of the pellets and it is taken into account during the analysis of the experimental data following the method explained in \cite{mennella1998} and based on the Bruggeman theory. More detailed treatment of dust coagulation by methods such as DDA have shown that it may increase the MAC by a factor of  two at most  \citep{kohler2011,mackowski2006,min2016}, that is, not enough to fully account for the discrepancy of the cosmic dust models and the experimental data. The enhancement of the emissivity of iron-rich analogues compared to the MAC of cosmic dust models commonly used for interpreting FIR/submm dust emission observations is related to the disordered nature of the samples, to the number, distribution, and nature of defects in their structure at microscopic scale and to the existence of absorption processes added to the classical Debye model \citep[we refer to][for more details]{demyk2017}. These additional absorption processes are more or less important depending on the structural state  of the material at the microscopic scale. \\

This study shows that the MAC of ferromagnesium silicates has the same qualitative behavior as the one of pure Mg-rich silicates in terms of dependence on the temperature and wavelength. The presence of iron oxides in the samples does not suppress this behavior, probably because the iron oxide phases are amorphous and also are not very abundant. This emphasizes the universality of this behavior in amorphous solids, whatever their composition, and the fact that it has to be taken into account in astronomical modeling. Indeed, any cosmic dust composed of amorphous silicates and/or oxides will be characterized by a MAC (i.e., an emissivity) that is lower at low temperature than at high temperature and which deviates from a single power law such as ${\lambda}^{-\beta}$. As for previous studies from \cite{coupeaud2011} and \cite{demyk2017}, the variation of the MAC of the Fe-rich analogues is observed for temperatures greater than 30 K. The MAC of the samples is identical in the 10 - 30 K range and then increases at 100, 200, and 300 K. In astronomical modeling, it is thus important to have a first guess of the dust temperature in order to use the MAC of the dust analogues measured at a temperature as close as possible to the dust temperature. In addition, considering the MAC averaged over all the samples and adopting the pessimistic assumption that coagulation effects are not properly taken into account (thus dividing the MAC by a factor of two), the value of <MAC$\mathrm{_{all}}$> at 10 K is two or three times higher than the modeled MAC at 1 mm and three or four times higher at 500 $\mu$m. The direct consequence of this is that cosmic dust models overestimate the dust mass compared to the use of an experimental MAC.

\section{Conclusions}

The MAC of eight ferromagnesium amorphous silicate analogues were measured in the MIR (5 - 40 $\mu$m) at room temperature and in the FIR/submm (30 $\mu$m - 1mm) at various temperatures (10, 30, 100, 200 and 300 K). The analogues are amorphous silicates of the mean composition of pyroxene with varying amounts of magnesium and iron: Mg$_{1-x}$Fe$_x$SiO$_3$ with $x$ = 0.1, 0.2, 0.3 and 0.4. Four samples were processed to modify the iron oxidation state within the materials leading to four additional samples having a modified structure and chemical homogeneity at microscopic scale compared to the non-processed samples. 

We find that the MAC of ferromagnesium amorphous silicates exhibits the same characteristics as other Mg-rich amorphous silicates. In the FIR, the MAC of the sample increases with the grain temperature as absorption processes are thermally activated. The wavelength at which the MAC changes depends on the samples and on the temperature in the range 100 - 200 $\mu$m. These thermal effects are observed above 30 K and we find that the MAC at 10 and 30K are identical for all samples. The overall spectral shape of the MAC differs from a power law in ${\lambda}^{-\beta}$, which is the usually adopted extrapolation in astronomical models. The value of $\beta$, defined as the local slope of the MAC, varies with the wavelength. For a given sample, and at a given wavelength, the value of $\beta$ is anti-correlated with the grain temperature. 

The qualitative agreement of the MAC of Fe-rich and Mg-rich amorphous silicates shows that more than the composition (Mg vs. Fe content), the structure at nanometer scale governs the FIR/submm absorption. Hence, any amorphous silicate grain, whatever its composition, should present complex behavior with wavelength and temperature. The modifications of the MAC induced by the processing do not show a common behavior in terms of shape and intensity. However, the comparison of the averaged MAC of the unprocessed samples with the one of the processed samples shows that they are different. We attribute these differences to an evolution of the amorphous silicate network of the samples during the processing. 

The MAC of the Fe-rich samples is much higher than the MAC in cosmic dust models. This is not due to compositional differences. We attribute this enhancement to absorption processes which are added to the Debye model. These absorption processes are characteristic of the amorphous nature of the dust and on the nature and distribution of defects of the disordered structure. This has important astronomical implications in terms of mass determination and elemental abundance constrains of cosmic dust models.


\begin{acknowledgements}
This work was supported by the French \emph{Agence National pour la recherche} project ANR-CIMMES and  by the Programme National PCMI of CNRS/INSU with INC/INP co-funded by CEA and CNES. We thank the referee, J. Nuth, for his comments that have helped to improve this manuscript. We thank A. Marra for providing us with the optical constant of hematite and V. Mennella for sharing its experimental data of the MAC of amorphous fayalite. We thank M. Roskosz for fruitful discussions about M\"ossbauer spectroscopy. 
\end{acknowledgements}

\bibliographystyle{aa} 
\bibliography{bib1} 

\begin{thebibliography}{50}
\expandafter\ifx\csname natexlab\endcsname\relax\def\natexlab#1{#1}\fi

\bibitem[{{Agladze} {et~al.}(1996){Agladze}, {Sievers}, {Jones}, {Burlitch}, \&
  {Beckwith}}]{agladze1996}
{Agladze}, N.~I., {Sievers}, A.~J., {Jones}, S.~A., {Burlitch}, J.~M., \&
  {Beckwith}, S.~V.~W. 1996, \apj, 462, 1026

\bibitem[{{Bohren} \& {Huffman}(1998)}]{bohren1998}
{Bohren}, C.~F. \& {Huffman}, D.~R. 1998, {Absorption and Scattering of Light
  by Small Particles}, ed. {Bohren, C.~F.~\& Huffman, D.~R.}

\bibitem[{{Bose} {et~al.}(2012){Bose}, {Floss}, {Stadermann}, {Stroud}, \&
  {Speck}}]{bose2012}
{Bose}, M., {Floss}, C., {Stadermann}, F.~J., {Stroud}, R.~M., \& {Speck},
  A.~K. 2012, \gca, 93, 77

\bibitem[{{Boudet} {et~al.}(2005){Boudet}, {Mutschke}, {Nayral}, {J{\"a}ger},
  {Bernard}, {Henning}, \& {Meny}}]{boudet2005}
{Boudet}, N., {Mutschke}, H., {Nayral}, C., {et~al.} 2005, \apj, 633, 272

\bibitem[{{Bradley}(1994)}]{bradley1994}
{Bradley}, J.~P. 1994, Science, 265, 925

\bibitem[{{Brubach} {et~al.}(2010){Brubach}, {Manceron}, {Rouzi{\`e}res},
  {Pirali}, {Balcon}, {Tchana}, {Boudon}, {Tudorie}, {Huet}, {Cuisset}, \&
  {Roy}}]{brubach2010}
{Brubach}, J., {Manceron}, L., {Rouzi{\`e}res}, M., {et~al.} 2010, in American
  Institute of Physics Conference Series, Vol. 1214, American Institute of
  Physics Conference Series, ed. {A.~Predoi-Cross \& B.~E.~Billinghurst},
  81--84

\bibitem[{{Compi{\`e}gne} {et~al.}(2011){Compi{\`e}gne}, {Verstraete}, {Jones},
  {Bernard}, {Boulanger}, {Flagey}, {Le Bourlot}, {Paradis}, \&
  {Ysard}}]{compiegne2011}
{Compi{\`e}gne}, M., {Verstraete}, L., {Jones}, A., {et~al.} 2011, \aap, 525,
  A103+

\bibitem[{{Coupeaud} {et~al.}(2011){Coupeaud}, {Demyk}, {Meny}, {Nayral},
  {Delpech}, {Leroux}, {Depecker}, {Creff}, {Brubach}, \& {Roy}}]{coupeaud2011}
{Coupeaud}, A., {Demyk}, K., {Meny}, C., {et~al.} 2011, \aap, 535, A124

\bibitem[{{Demyk} {et~al.}(2013){Demyk}, {Meny}, {Leroux}, {Depecker},
  {Nayral}, {Delpech}, {Ojo}, {Brubach}, \& {Roy}}]{demyk2013}
{Demyk}, K., {Meny}, C., {Leroux}, H., {et~al.} 2013, in PoS(LDCU2013)044

\bibitem[{{Demyk} {et~al.}(2017){Demyk}, {Meny}, {Lu}, {Papatheodorou},
  {Toplis}, {Leroux}, {Depecker}, {Brubach}, {Roy}, {Nayral}, {Ojo}, {Delpech},
  {Paradis}, \& {Gromov}}]{demyk2017}
{Demyk}, K., {Meny}, C., {Lu}, X.-H., {et~al.} 2017, \aap, 600, A123

\bibitem[{{Dorschner} {et~al.}(1995){Dorschner}, {Begemann}, {Henning},
  {Jaeger}, \& {Mutschke}}]{dorschner1995}
{Dorschner}, J., {Begemann}, B., {Henning}, T., {Jaeger}, C., \& {Mutschke}, H.
  1995, \aap, 300, 503

\bibitem[{{Draine} \& {Flatau}(2013)}]{draine2013}
{Draine}, B.~T. \& {Flatau}, P.~J. 2013, ArXiv e-prints
  [\eprint[arXiv]{1305.6497}]

\bibitem[{{Draine} \& {Hensley}(2013)}]{draine2013a}
{Draine}, B.~T. \& {Hensley}, B. 2013, \apj, 765, 159

\bibitem[{{Draine} \& {Lee}(1984)}]{draine1984}
{Draine}, B.~T. \& {Lee}, H.~M. 1984, \apj, 285, 89

\bibitem[{{Dwek}(2016)}]{dwek2016}
{Dwek}, E. 2016, \apj, 825, 136

\bibitem[{{Ferreira da Silva} {et~al.}(1992){Ferreira da Silva}, {Waerenborgh},
  {Navarro}, \& {Cabral}}]{ferreiradasilva1992}
{Ferreira da Silva}, M., G., {Waerenborgh}, J., C., {Navarro}, J., \& {Cabral},
  J. 1992, J. Non-Cryst. Solids, 147 \& 143, 146

\bibitem[{{Galliano} {et~al.}(2011){Galliano}, {Hony}, {Bernard}, {Bot},
  {Madden}, {Roman-Duval}, {Galametz}, {Li}, {Meixner}, {Engelbracht},
  {Lebouteiller}, {Misselt}, {Montiel}, {Panuzzo}, {Reach}, \&
  {Skibba}}]{galliano2011}
{Galliano}, F., {Hony}, S., {Bernard}, J.-P., {et~al.} 2011, \aap, 536, A88

\bibitem[{{Gillot} {et~al.}(2009){Gillot}, {Roskosz}, {Depecker}, {Roussel}, \&
  {Leroux}}]{gillot2009}
{Gillot}, J., {Roskosz}, M., {Depecker}, C., {Roussel}, P., \& {Leroux}, H.
  2009, in Lunar and Planetary Institute Science Conference Abstracts, Vol.~40,
  Lunar and Planetary Institute Science Conference Abstracts, 1763--+

\bibitem[{Gurevich {et~al.}(2003)Gurevich, Parshin, \& Schober}]{gurevich2003}
Gurevich, V.~L., Parshin, D.~A., \& Schober, H.~R. 2003, Phys. Rev. B, 67,
  094203

\bibitem[{{Henning} {et~al.}(1995){Henning}, {Michel}, \&
  {Stognienko}}]{henning1995}
{Henning}, T., {Michel}, B., \& {Stognienko}, R. 1995, \planss, 43, 1333

\bibitem[{{Henning} \& {Mutschke}(1997)}]{henning1997}
{Henning}, T. \& {Mutschke}, H. 1997, \aap, 327, 743

\bibitem[{{Jayasuriya} {et~al.}(2004){Jayasuriya}, {O'Neill}, {Berry}, \&
  {Campbell}}]{jayasuriya2004}
{Jayasuriya}, k., {O'Neill}, H., {Berry}, A., \& {Campbell}, S. 2004, Am. Min.,
  89, 1597

\bibitem[{{Jenkins}(2009)}]{jenkins2009}
{Jenkins}, E.~B. 2009, \apj, 700, 1299

\bibitem[{{Jones} {et~al.}(2013){Jones}, {Fanciullo}, {K{\"o}hler},
  {Verstraete}, {Guillet}, {Bocchio}, \& {Ysard}}]{jones2013}
{Jones}, A.~P., {Fanciullo}, L., {K{\"o}hler}, M., {et~al.} 2013, \aap, 558,
  A62

\bibitem[{{Juvela} {et~al.}(2015){Juvela}, {Demyk}, {Doi}, {Hughes},
  {Lef{\`e}vre}, {Marshall}, {Meny}, {Montillaud}, {Pagani}, {Paradis},
  {Ristorcelli}, {Malinen}, {Montier}, {Paladini}, {Pelkonen}, \&
  {Rivera-Ingraham}}]{juvela2015}
{Juvela}, M., {Demyk}, K., {Doi}, Y., {et~al.} 2015, \aap, 584, A94

\bibitem[{{Juvela} {et~al.}(2013){Juvela}, {Montillaud}, {Ysard}, \&
  {Lunttila}}]{juvela2013}
{Juvela}, M., {Montillaud}, J., {Ysard}, N., \& {Lunttila}, T. 2013, \aap, 556,
  A63

\bibitem[{{Juvela} \& {Ysard}(2012{\natexlab{a}})}]{juvela2012a}
{Juvela}, M. \& {Ysard}, N. 2012{\natexlab{a}}, \aap, 541, A33

\bibitem[{{Juvela} \& {Ysard}(2012{\natexlab{b}})}]{juvela2012b}
{Juvela}, M. \& {Ysard}, N. 2012{\natexlab{b}}, \aap, 539, A71

\bibitem[{{K{\"o}hler} {et~al.}(2011){K{\"o}hler}, {Guillet}, \&
  {Jones}}]{kohler2011}
{K{\"o}hler}, M., {Guillet}, V., \& {Jones}, A. 2011, \aap, 528, A96+

\bibitem[{{K{\"o}hler} {et~al.}(2014){K{\"o}hler}, {Jones}, \&
  {Ysard}}]{koehler2014}
{K{\"o}hler}, M., {Jones}, A., \& {Ysard}, N. 2014, \aap, 565, L9

\bibitem[{{K{\"o}hler} {et~al.}(2012){K{\"o}hler}, {Stepnik}, {Jones},
  {Guillet}, {Abergel}, {Ristorcelli}, \& {Bernard}}]{koehler2012}
{K{\"o}hler}, M., {Stepnik}, B., {Jones}, A.~P., {et~al.} 2012, \aap, 548, A61

\bibitem[{{Li} \& {Draine}(2001)}]{li2001}
{Li}, A. \& {Draine}, B.~T. 2001, \apj, 554, 778

\bibitem[{{Lyubutin} {et~al.}(2009){Lyubutin}, R., {Korzhetskiy}, V., \&
  K.}]{lyubutin2009}
{Lyubutin}, I.~S., R., L.~C., {Korzhetskiy}, V., V., D.~T., \& K., C.~R. 2009,
  Applied Optics, 106

\bibitem[{Mackowski(2006)}]{mackowski2006}
Mackowski, D.~W. 2006, Journal of Quantitative Spectroscopy and Radiative
  Transfer, 100, 237

\bibitem[{{Malinen} {et~al.}(2011){Malinen}, {Juvela}, {Collins}, {Lunttila},
  \& {Padoan}}]{malinen2011}
{Malinen}, J., {Juvela}, M., {Collins}, D.~C., {Lunttila}, T., \& {Padoan}, P.
  2011, \aap, 530, A101

\bibitem[{{Marra} {et~al.}(2011){Marra}, {Lane}, {Orofino}, {Blanco}, \&
  {Fonti}}]{marra2011}
{Marra}, A.~C., {Lane}, M.~D., {Orofino}, V., {Blanco}, A., \& {Fonti}, S.
  2011, \icarus, 211, 839

\bibitem[{{Meisner} \& {Finkbeiner}(2015)}]{meisner2015}
{Meisner}, A.~M. \& {Finkbeiner}, D.~P. 2015, \apj, 798, 88

\bibitem[{{Mennella} {et~al.}(1998){Mennella}, {Brucato}, {Colangeli},
  {Palumbo}, {Rotundi}, \& {Bussoletti}}]{mennella1998}
{Mennella}, V., {Brucato}, J.~R., {Colangeli}, L., {et~al.} 1998, \apj, 496,
  1058

\bibitem[{{Meny} {et~al.}(2007){Meny}, {Gromov}, {Boudet}, {Bernard},
  {Paradis}, \& {Nayral}}]{meny2007}
{Meny}, C., {Gromov}, V., {Boudet}, N., {et~al.} 2007, \aap, 468, 171

\bibitem[{{Min} {et~al.}(2016){Min}, {Rab}, {Woitke}, {Dominik}, \&
  {M{\'e}nard}}]{min2016}
{Min}, M., {Rab}, C., {Woitke}, P., {Dominik}, C., \& {M{\'e}nard}, F. 2016,
  \aap, 585, A13

\bibitem[{{Mohr} {et~al.}(2013){Mohr}, {Mutschke}, \& {Lewen}}]{mohr2013}
{Mohr}, P., {Mutschke}, H., \& {Lewen}, F. 2013, in PoS(LDCU2013)140

\bibitem[{{Paradis} {et~al.}(2009){Paradis}, {Bernard}, \&
  {M{\'e}ny}}]{paradis2009}
{Paradis}, D., {Bernard}, J., \& {M{\'e}ny}, C. 2009, \aap, 506, 745

\bibitem[{{Paradis} {et~al.}(2014){Paradis}, {M{\'e}ny}, {Noriega-Crespo},
  {Paladini}, {Bernard}, {Bot}, {Cambr{\'e}sy}, {Demyk}, {Gromov},
  {Rivera-Ingraham}, \& {Veneziani}}]{paradis2014}
{Paradis}, D., {M{\'e}ny}, C., {Noriega-Crespo}, A., {et~al.} 2014, ArXiv
  e-prints [\eprint[arXiv]{1409.6892}]

\bibitem[{{Planck Collaboration} {et~al.}(2014{\natexlab{a}}){Planck
  Collaboration}, {Abergel}, {Ade}, {Aghanim}, {Alves}, {Aniano},
  {Armitage-Caplan}, {Arnaud}, {Ashdown}, {Atrio-Barandela}, \&
  et~al.}]{planck2013-XI}
{Planck Collaboration}, {Abergel}, A., {Ade}, P.~A.~R., {et~al.}
  2014{\natexlab{a}}, \aap, 571, A11

\bibitem[{{Planck Collaboration} {et~al.}(2014{\natexlab{b}}){Planck
  Collaboration}, {Abergel}, {Ade}, {Aghanim}, {Alves}, {Aniano}, {Arnaud},
  {Ashdown}, {Aumont}, {Baccigalupi}, {Banday}, {Barreiro}, {Bartlett},
  {Battaner}, {Benabed}, {Benoit-L{\'e}vy}, {Bernard}, {Bersanelli},
  {Bielewicz}, {Bobin}, {Bonaldi}, {Bond}, {Bouchet}, {Boulanger}, {Burigana},
  {Cardoso}, {Catalano}, {Chamballu}, {Chiang}, {Christensen}, {Clements},
  {Colombi}, {Colombo}, {Couchot}, {Crill}, {Cuttaia}, {Danese}, {Davis}, {de
  Bernardis}, {de Rosa}, {de Zotti}, {Delabrouille}, {D{\'e}sert}, {Dickinson},
  {Diego}, {Dole}, {Donzelli}, {Dor{\'e}}, {Douspis}, {Dupac}, {Efstathiou},
  {En{\ss}lin}, {Eriksen}, {Falgarone}, {Finelli}, {Forni}, {Frailis},
  {Franceschi}, {Galeotta}, {Ganga}, {Ghosh}, {Giard}, {Giraud-H{\'e}raud},
  {Gonz{\'a}lez-Nuevo}, {G{\'o}rski}, {Gregorio}, {Gruppuso}, {Guillet},
  {Hansen}, {Harrison}, {Helou}, {Henrot-Versill{\'e}},
  {Hern{\'a}ndez-Monteagudo}, {Herranz}, {Hildebrandt}, {Hivon}, {Hobson},
  {Holmes}, {Hornstrup}, {Hovest}, {Huffenberger}, {Jaffe}, {Jaffe}, {Joncas},
  {Jones}, {Jones}, {Juvela}, {Kalberla}, {Keih{\"a}nen}, {Kerp}, {Keskitalo},
  {Kisner}, {Kneissl}, {Knoche}, {Kunz}, {Kurki-Suonio}, {Lagache},
  {L{\"a}hteenm{\"a}ki}, {Lamarre}, {Lasenby}, {Lawrence}, {Leonardi},
  {Levrier}, {Liguori}, {Lilje}, {Linden-V{\o}rnle}, {L{\'o}pez-Caniego},
  {Lubin}, {Mac{\'{\i}}as-P{\'e}rez}, {Maffei}, {Maino}, {Mandolesi}, {Maris},
  {Marshall}, {Martin}, {Mart{\'{\i}}nez-Gonz{\'a}lez}, {Masi}, {Massardi},
  {Matarrese}, {Mazzotta}, {Melchiorri}, {Mendes}, {Mennella}, {Migliaccio},
  {Mitra}, {Miville-Desch{\^e}nes}, {Moneti}, {Montier}, {Morgante},
  {Mortlock}, {Munshi}, {Murphy}, {Naselsky}, {Nati}, {Natoli}, {Noviello},
  {Novikov}, {Novikov}, {Oxborrow}, {Pagano}, {Pajot}, {Paoletti}, {Pasian},
  {Perdereau}, {Perotto}, {Perrotta}, {Piacentini}, {Piat}, {Pierpaoli},
  {Pietrobon}, {Plaszczynski}, {Pointecouteau}, {Polenta}, {Ponthieu}, {Popa},
  {Pratt}, {Prunet}, {Puget}, {Rachen}, {Reach}, {Rebolo}, {Reinecke},
  {Remazeilles}, {Renault}, {Ricciardi}, {Riller}, {Ristorcelli}, {Rocha},
  {Rosset}, {Roudier}, {Rusholme}, {Sandri}, {Savini}, {Spencer}, {Starck},
  {Sureau}, {Sutton}, {Suur-Uski}, {Sygnet}, {Tauber}, {Terenzi}, {Toffolatti},
  {Tomasi}, {Tristram}, {Tucci}, {Umana}, {Valenziano}, {Valiviita}, {Van
  Tent}, {Verstraete}, {Vielva}, {Villa}, {Wade}, {Wandelt}, {Winkel}, {Yvon},
  {Zacchei}, \& {Zonca}}]{planck-XVII-Int-2014}
{Planck Collaboration}, {Abergel}, A., {Ade}, P.~A.~R., {et~al.}
  2014{\natexlab{b}}, \aap, 566, A55

\bibitem[{{Planck Collaboration} {et~al.}(2011{\natexlab{a}}){Planck
  Collaboration}, {Abergel}, {Ade}, {Aghanim}, {Arnaud}, {Ashdown}, {Aumont},
  {Baccigalupi}, {Balbi}, {Banday}, {Barreiro}, {Bartlett}, {Battaner},
  {Benabed}, {Beno{\^i}t}, {Bernard}, {Bersanelli}, {Bhatia}, {Bock},
  {Bonaldi}, {Bond}, {Borrill}, {Bouchet}, {Boulanger}, {Bucher}, {Burigana},
  {Cabella}, {Cardoso}, {Catalano}, {Cay{\'o}n}, {Challinor}, {Chamballu},
  {Chiang}, {Chiang}, {Christensen}, {Clements}, {Colombi}, {Couchot},
  {Coulais}, {Crill}, {Cuttaia}, {Danese}, {Davies}, {Davis}, {de Bernardis},
  {de Gasperis}, {de Rosa}, {de Zotti}, {Delabrouille}, {Delouis},
  {D{\'e}sert}, {Dickinson}, {Dobashi}, {Donzelli}, {Dor{\'e}}, {D{\"o}rl},
  {Douspis}, {Dupac}, {Efstathiou}, {En{\ss}lin}, {Eriksen}, {Finelli},
  {Forni}, {Frailis}, {Franceschi}, {Galeotta}, {Ganga}, {Giard}, {Giardino},
  {Giraud-H{\'e}raud}, {Gonz{\'a}lez-Nuevo}, {G{\'o}rski}, {Gratton},
  {Gregorio}, {Gruppuso}, {Guillet}, {Hansen}, {Harrison},
  {Henrot-Versill{\'e}}, {Herranz}, {Hildebrandt}, {Hivon}, {Hobson}, {Holmes},
  {Hovest}, {Hoyland}, {Huffenberger}, {Jaffe}, {Jones}, {Jones}, {Juvela},
  {Keih{\"a}nen}, {Keskitalo}, {Kisner}, {Kneissl}, {Knox}, {Kurki-Suonio},
  {Lagache}, {Lamarre}, {Lasenby}, {Laureijs}, {Lawrence}, {Leach}, {Leonardi},
  {Leroy}, {Linden-V{\o}rnle}, {L{\'o}pez-Caniego}, {Lubin},
  {Mac{\'{\i}}as-P{\'e}rez}, {MacTavish}, {Maffei}, {Mandolesi}, {Mann},
  {Maris}, {Marshall}, {Martin}, {Mart{\'{\i}}nez-Gonz{\'a}lez}, {Masi},
  {Matarrese}, {Matthai}, {Mazzotta}, {McGehee}, {Meinhold}, {Melchiorri},
  {Mendes}, {Mennella}, {Mitra}, {Miville-Desch{\^e}nes}, {Moneti}, {Montier},
  {Morgante}, {Mortlock}, {Munshi}, {Murphy}, {Naselsky}, {Natoli},
  {Netterfield}, {N{\o}rgaard-Nielsen}, {Noviello}, {Novikov}, {Novikov},
  {Osborne}, {Pajot}, {Paladini}, {Pasian}, {Patanchon}, {Perdereau},
  {Perotto}, {Perrotta}, {Piacentini}, {Piat}, {Plaszczynski}, {Pointecouteau},
  {Polenta}, {Ponthieu}, {Poutanen}, {Pr{\'e}zeau}, {Prunet}, {Puget}, {Reach},
  {Rebolo}, {Reinecke}, {Renault}, {Ricciardi}, {Riller}, {Ristorcelli},
  {Rocha}, {Rosset}, {Rubi{\~n}o-Mart{\'{\i}}n}, {Rusholme}, {Sandri},
  {Santos}, {Savini}, {Scott}, {Seiffert}, {Shellard}, {Smoot}, {Starck},
  {Stivoli}, {Stolyarov}, {Sudiwala}, {Sygnet}, {Tauber}, {Terenzi},
  {Toffolatti}, {Tomasi}, {Torre}, {Tristram}, {Tuovinen}, {Umana},
  {Valenziano}, {Verstraete}, {Vielva}, {Villa}, {Vittorio}, {Wade}, {Wandelt},
  {Yvon}, {Zacchei}, \& {Zonca}}]{planck2011_early_XXV}
{Planck Collaboration}, {Abergel}, A., {Ade}, P.~A.~R., {et~al.}
  2011{\natexlab{a}}, \aap, 536, A25

\bibitem[{{Planck Collaboration} {et~al.}(2011{\natexlab{b}}){Planck
  Collaboration}, {Ade}, {Aghanim}, {Arnaud}, {Ashdown}, {Aumont},
  {Baccigalupi}, {Balbi}, {Banday}, {Barreiro}, \&
  et~al.}]{planck2011_early_XXIII}
{Planck Collaboration}, {Ade}, P.~A.~R., {Aghanim}, N., {et~al.}
  2011{\natexlab{b}}, \aap, 536, A23

\bibitem[{{Richey} {et~al.}(2013){Richey}, {Kinzer}, {Cataldo}, {Wollack},
  {Nuth}, {Benford}, {Silverberg}, \& {Rinehart}}]{richey2013}
{Richey}, C.~R., {Kinzer}, R.~E., {Cataldo}, G., {et~al.} 2013, \apj, 770, 46

\bibitem[{{Scott} \& {Duley}(1996)}]{scottduley1996}
{Scott}, A. \& {Duley}, W.~W. 1996, \apjs, 105, 401

\bibitem[{{Shetty} {et~al.}(2009){Shetty}, {Kauffmann}, {Schnee}, {Goodman}, \&
  {Ercolano}}]{shetty2009b}
{Shetty}, R., {Kauffmann}, J., {Schnee}, S., {Goodman}, A.~A., \& {Ercolano},
  B. 2009, \apj, 696, 2234

\end{thebibliography}

\appendix
\section{Comparison with astronomical models.}
\label{astro_models}
We present here the comparison of the MAC of each sample with the MAC calculated for the "astrosilicate model" \citep{li2001} and for the "Themis" model \citep{jones2013}. The calculations are performed using Mie theory \citep{bohren1998} for a spherical particle of size of 100 nm, for spherical grain populations with a log-normal size distribution centred at 1 $\mu$m, and for a continuous distribution of ellipsoids (CDE). For the spheroidal (prolate grain with an axis ratio of 2) grain population (with a log-normal size distribution centred at 1 $\mu$m), we have used the DDA code DDSCAT 7.3 developed by \citet{draine2013} to calculate the MAC. 

\begin{figure}[!h]
\begin{center}
\includegraphics[scale=.38,trim={1cm 1cm 0cm 1cm},clip]{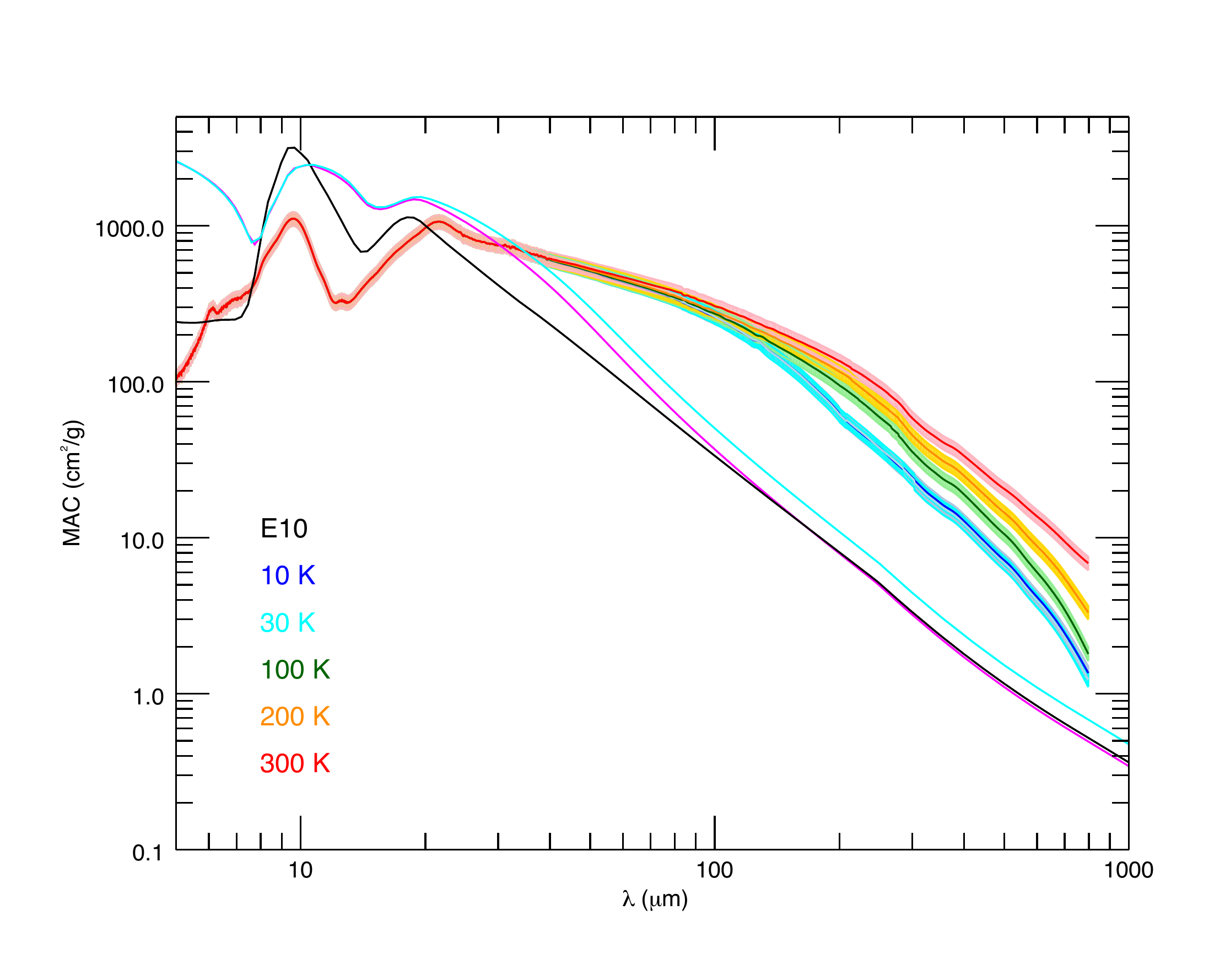}
  \end{center}
       \caption{Comparison of the MAC of sample E10 with the MAC calculated from astronomical dust models. The MAC calculated with  the "astrosilicates" is shown for a 0.1 $\mu$m size grain (black), for a log-normal grain size distribution with a mean diameter of 1 $\mu$m for spherical grains (magenta), and for prolate grains (cyan) and for a continuous distribution of ellipsoids (CDE, green). The MAC calculated with the "Themis" dust model is shown for a spherical grain of 100 nm in diameter for amorphous forsterite (purple) and amorphous enstatite (orange).}
         \label{comp_model_E10}
 \end{figure}

\begin{figure}[!h]
\begin{center}
\includegraphics[scale=.31,trim={1cm 1cm 0cm 1cm},clip]{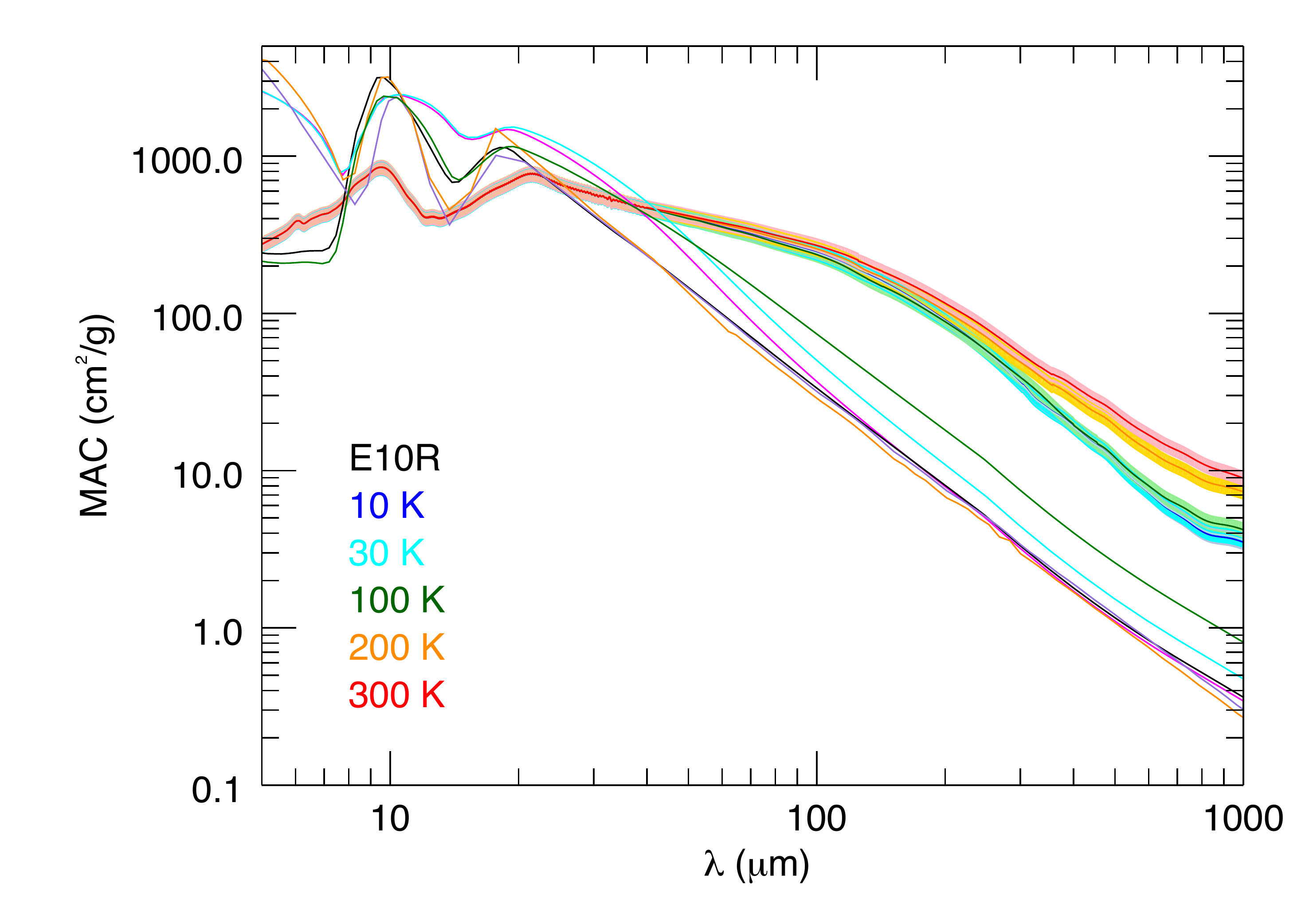}
  \end{center}
       \caption{Comparison of the MAC of sample E10R with the MAC calculated from astronomical dust models. The MAC calculated with  the "astrosilicates" is shown for a 0.1 $\mu$m size grain (black), for a log-normal grain size distribution with a mean diameter of 1 $\mu$m for spherical grains (magenta) and prolate grains (cyan), and for a continuous distribution of ellipsoids (CDE, green). The MAC calculated with the "Themis" dust model is shown for a spherical grain of 100 nm in diameter for amorphous forsterite (purple) and amorphous enstatite (orange).}
         \label{comp_model_E10R}
 \end{figure}

\begin{figure}[!t]
\begin{center}
\includegraphics[scale=.31,trim={1cm 1cm 0cm 1cm},clip]{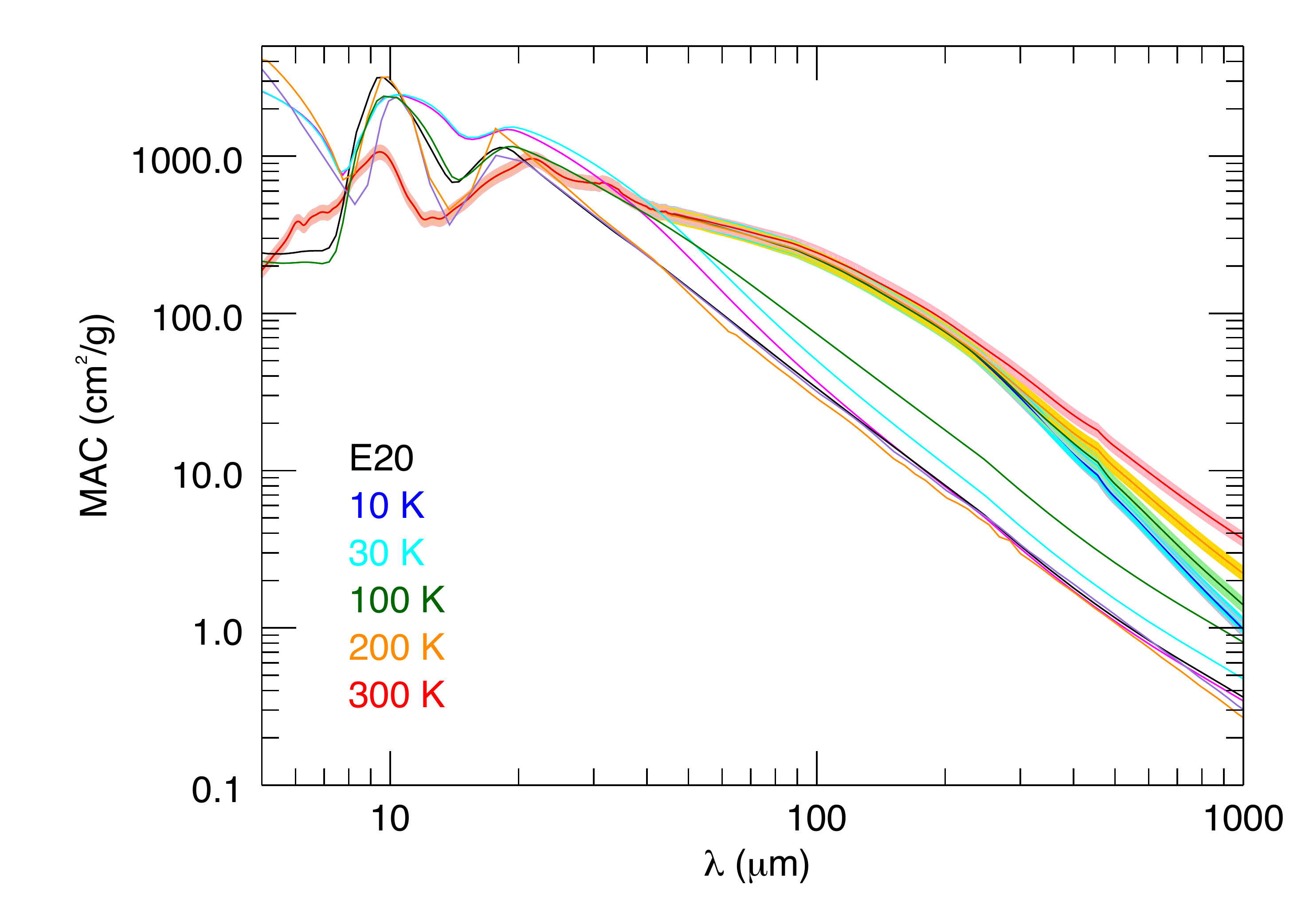}
  \end{center}
       \caption{Comparison of the MAC of sample E20 with the MAC calculated from astronomical dust models. The MAC calculated with  the "astrosilicates" is shown for a 0.1 $\mu$m size grain (black), for a log-normal grain size distribution with a mean diameter of 1 $\mu$m for spherical grains (magenta) and prolate grains (cyan), and for a continuous distribution of ellipsoids (CDE, green). The MAC calculated with the "Themis" dust model is shown for a spherical grain of 100 nm in diameter for amorphous forsterite (purple) and amorphous enstatite (orange).}
         \label{comp_model_E20}
 \end{figure}

\begin{figure}[!t]
\begin{center}
\includegraphics[scale=.31,trim={1cm 1cm 0cm 1cm},clip]{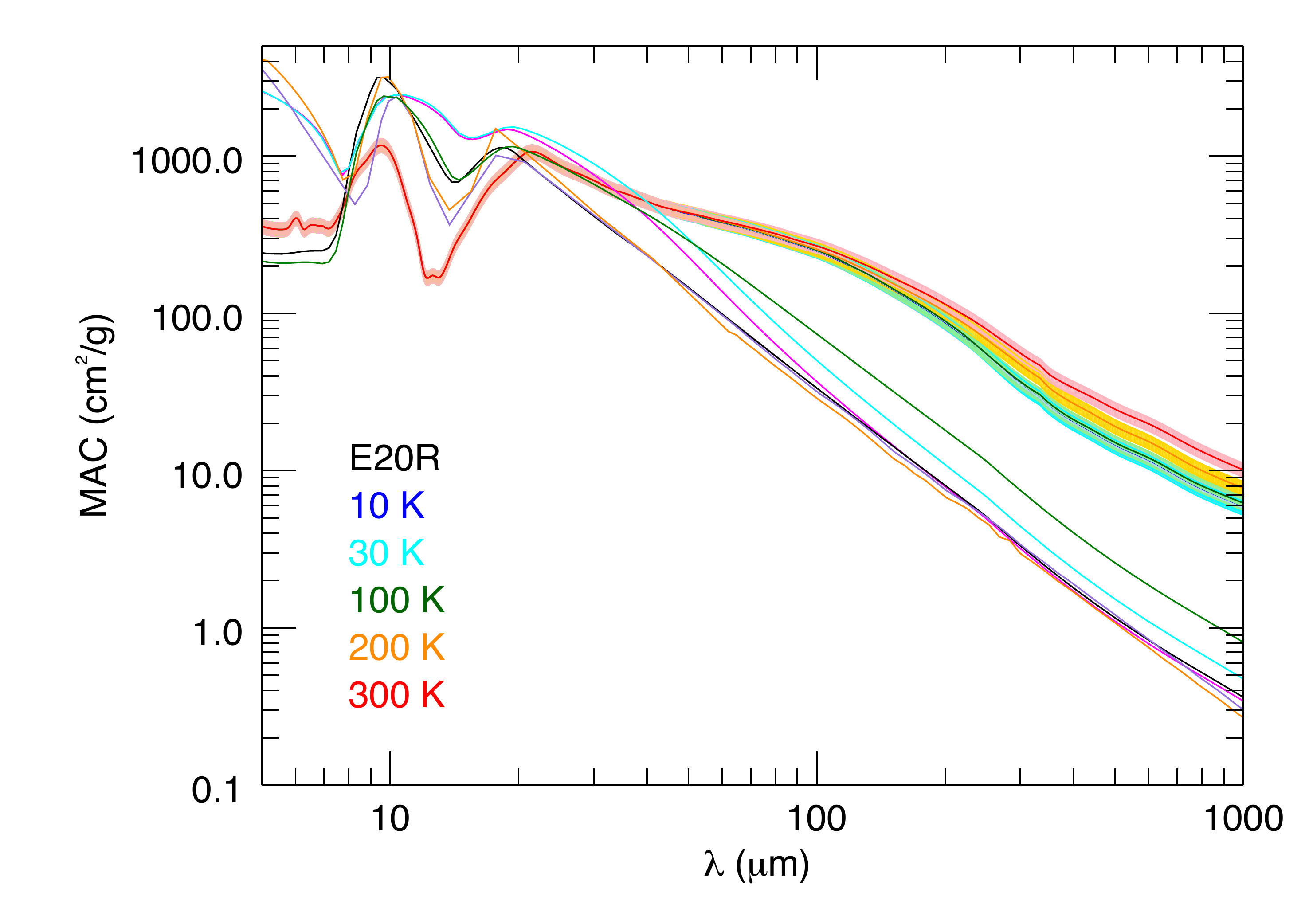}\\
  \end{center}
       \caption{Comparison of the MAC of sample E20R with the MAC calculated from astronomical dust models. The MAC calculated with  the "astrosilicates" is shown for a 0.1 $\mu$m size grain (black), for a log-normal grain size distribution with a mean diameter of 1 $\mu$m for spherical grains (magenta) and prolate grains (cyan), and for a continuous distribution of ellipsoids (CDE, green). The MAC calculated with the "Themis" dust model is shown for a spherical grain of 100 nm in diameter for amorphous forsterite (purple) and amorphous enstatite (orange).}
         \label{comp_model_E20R}
 \end{figure}

\begin{figure}[!t]
\begin{center}
\includegraphics[scale=.31,trim={1cm 1cm 0cm 1cm},clip]{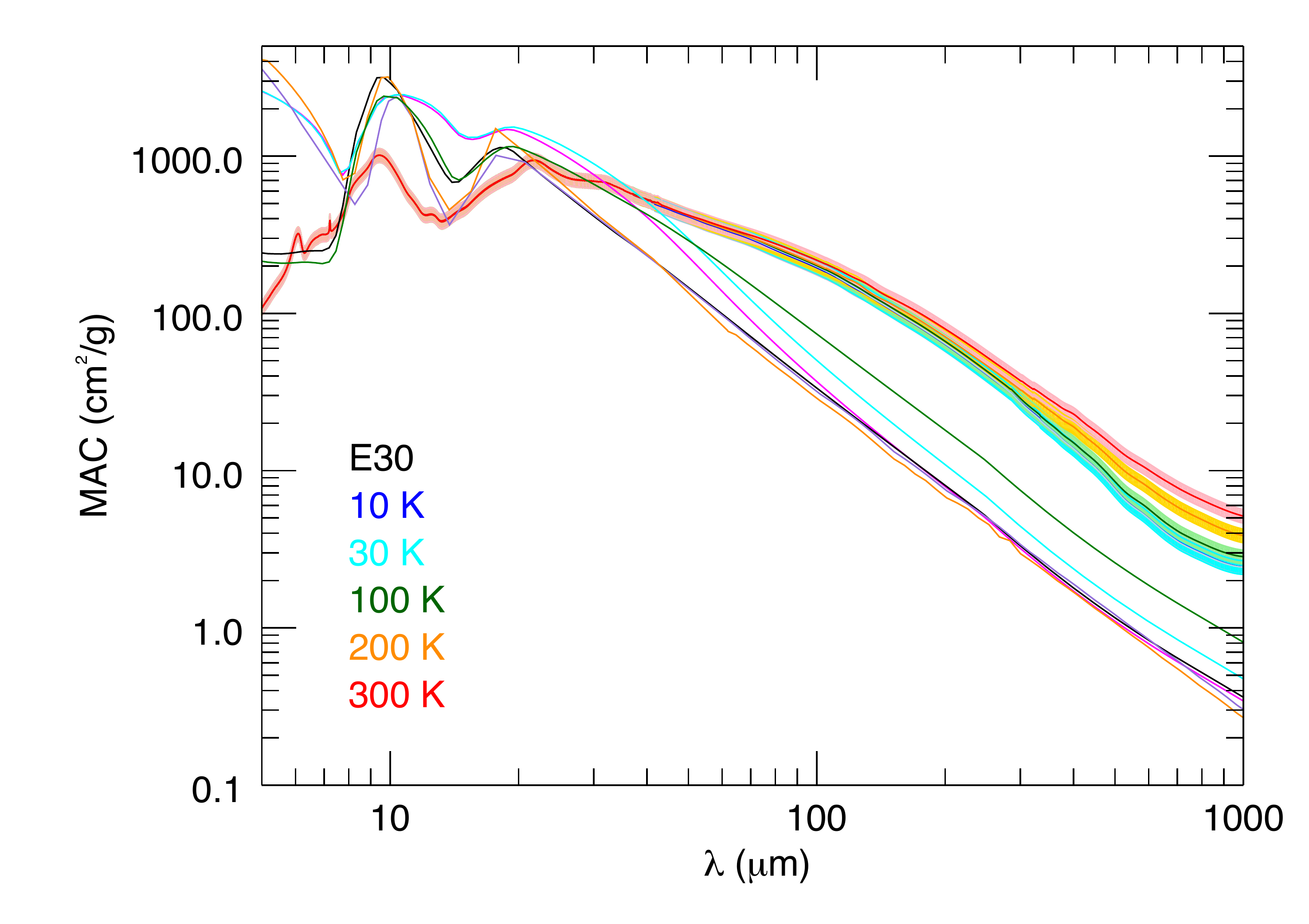}
 \end{center}
        \caption{Comparison of the MAC of sample E30 with the MAC calculated from astronomical dust models. The MAC calculated with  the "astrosilicates" is shown for a 0.1 $\mu$m size grain (black), for a log-normal grain size distribution with a mean diameter of 1 $\mu$m for spherical grains (magenta) and prolate grains (cyan), and for a continuous distribution of ellipsoids (CDE, green). The MAC calculated with the "Themis" dust model is shown for a spherical grain of 100 nm in diameter for amorphous forsterite (purple) and amorphous enstatite (orange).}
                 \label{comp_model_E30}
 \end{figure}

\begin{figure}[!t]
\begin{center}
\includegraphics[scale=.31,trim={1cm 1cm 0cm 1cm},clip]{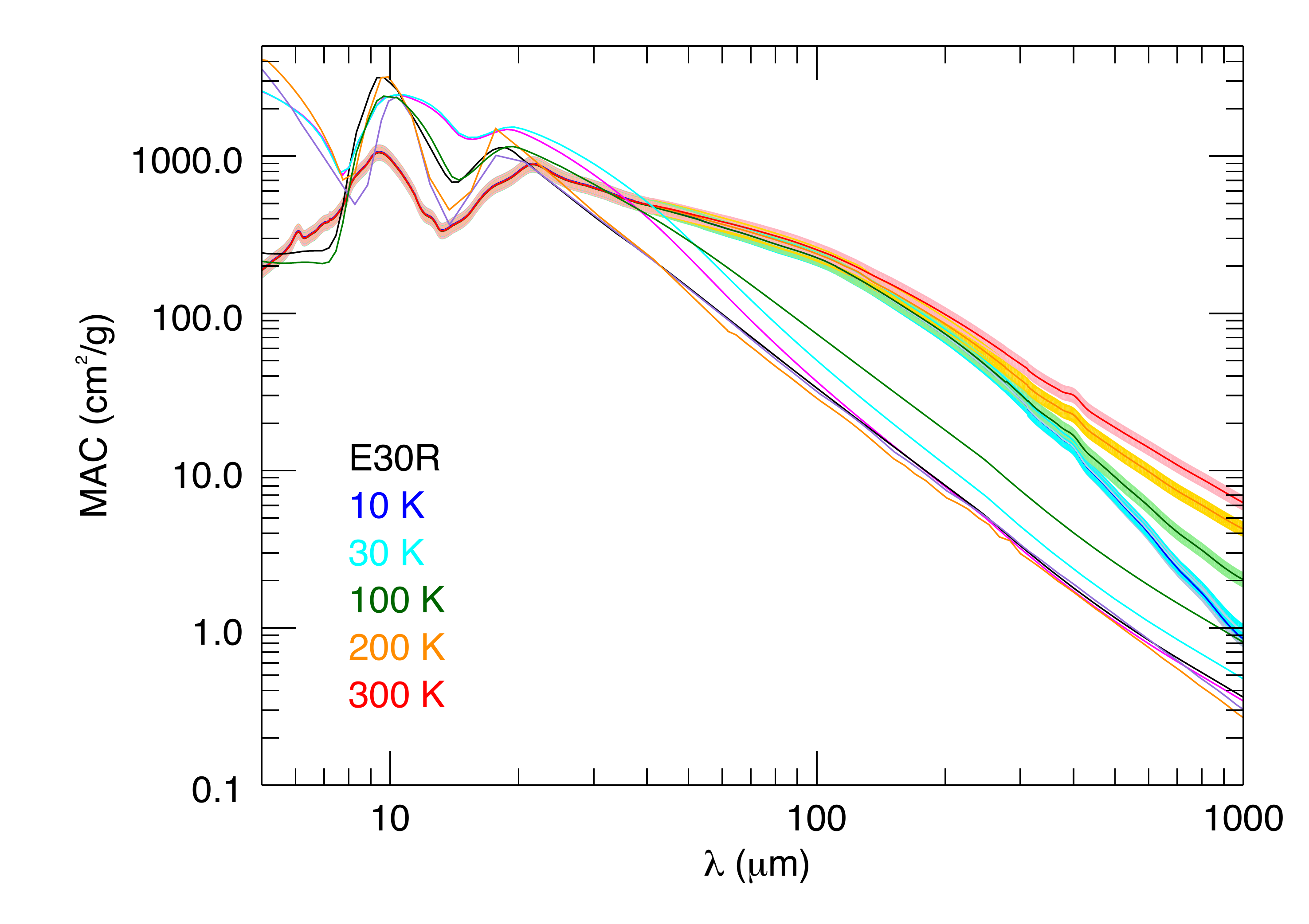}\\
 \end{center}
        \caption{Comparison of the MAC of sample E30R with the MAC calculated from astronomical dust models. The MAC calculated with  the "astrosilicates" is shown for a 0.1 $\mu$m size grain (black), for a log-normal grain size distribution with a mean diameter of 1 $\mu$m for spherical grains (magenta) and prolate grains (cyan), and for a continuous distribution of ellipsoids (CDE, green). The MAC calculated with the "Themis" dust model is shown for a spherical grain of 100 nm in diameter for amorphous forsterite (purple) and amorphous enstatite (orange).}
         \label{comp_model_E30R}
 \end{figure}

\begin{figure}[!t]
\begin{center}
\includegraphics[scale=.31,trim={1cm 1cm 0cm 1cm},clip]{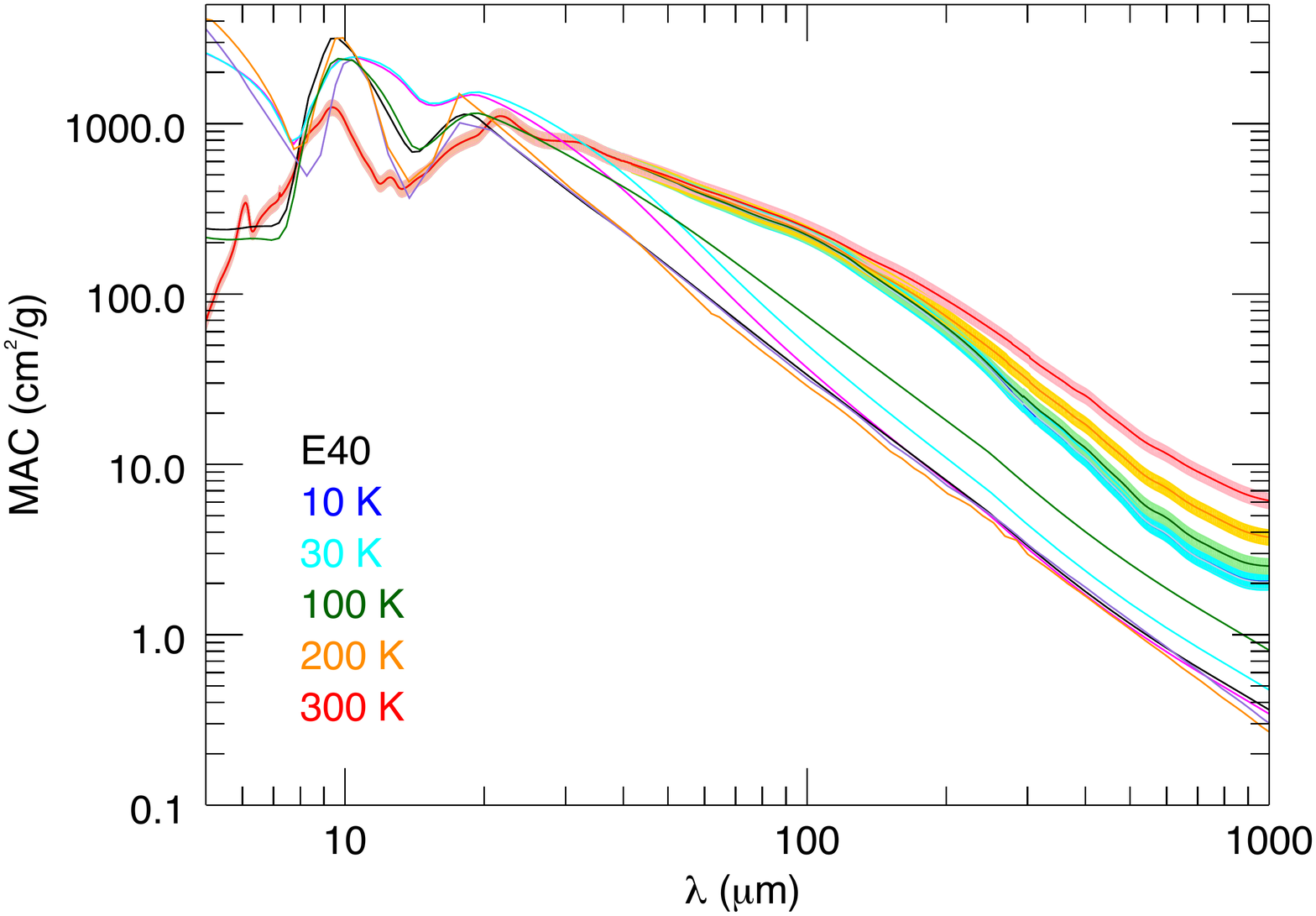}
 \end{center}
        \caption{Comparison of the MAC of sample E40 with the MAC calculated from astronomical dust models. The MAC calculated with  the "astrosilicates" is shown for a 0.1 $\mu$m size grain (black), for a log-normal grain size distribution with a mean diameter of 1 $\mu$m for spherical grains (magenta) and prolate grains (cyan), and for a continuous distribution of ellipsoids (CDE, green). The MAC calculated with the "Themis" dust model is shown for a spherical grain of 100 nm in diameter for amorphous forsterite (purple) and amorphous enstatite (orange).}
         \label{comp_model_E40}
 \end{figure}

\begin{figure}[!t]
\begin{center}
\includegraphics[scale=.31,trim={1cm 1cm 0cm 1cm},clip]{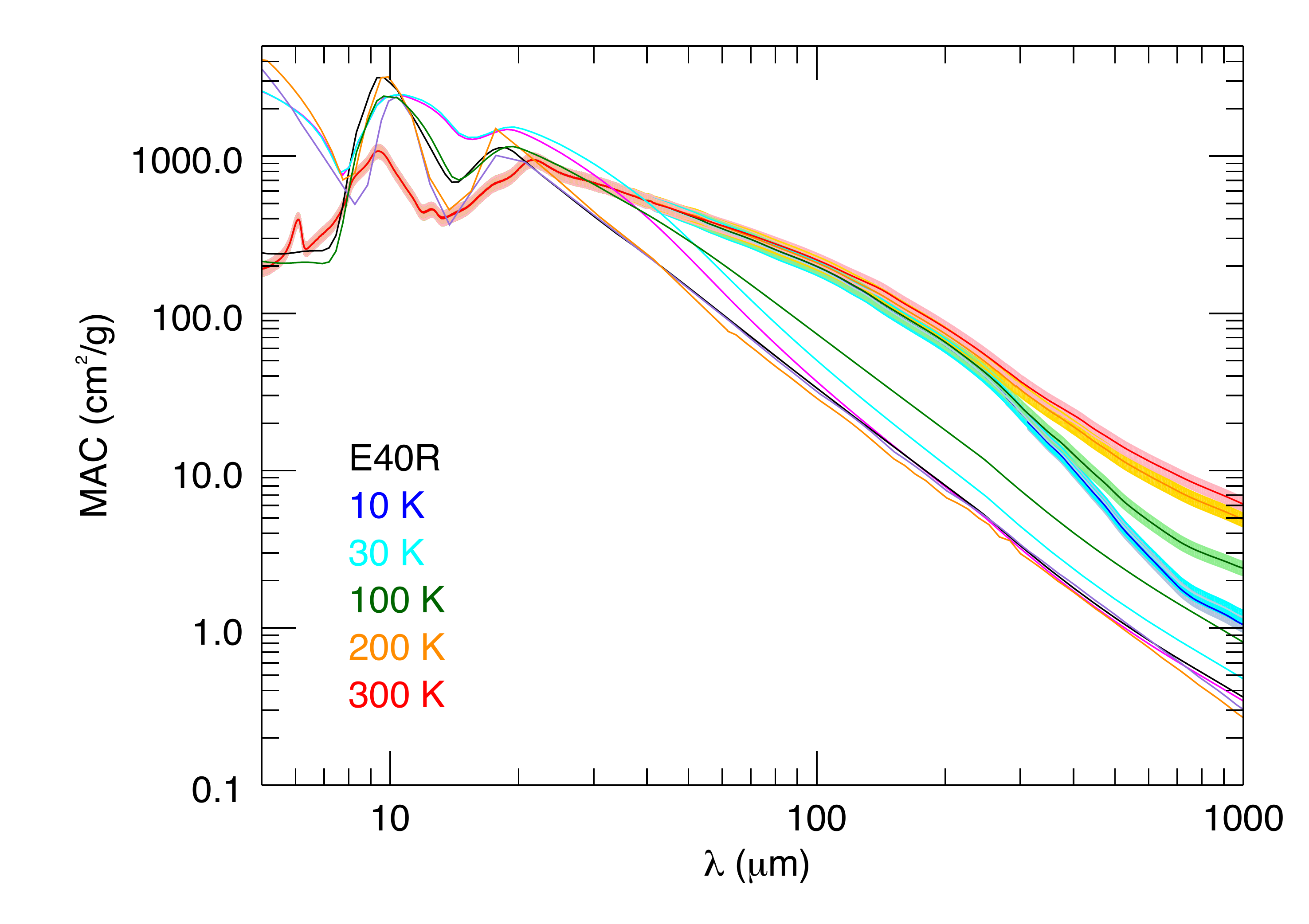}
 \end{center}
         \caption{Comparison of the MAC of sample E40R with the MAC calculated from astronomical dust models. The MAC calculated with  the "astrosilicates" is shown for a 0.1 $\mu$m size grain (black), for a log-normal grain size distribution with a mean diameter of 1 $\mu$m for spherical grains (magenta) and prolate grains (cyan), and for a continuous distribution of ellipsoids (CDE, green). The MAC calculated with the "Themis" dust model is shown for a spherical grain of 100 nm in diameter for amorphous forsterite (purple) and amorphous enstatite (orange).}
         \label{comp_model_E40R}
 \end{figure}

\end{document}